\documentclass[aps,preprint,amsmath,amssymb]{revtex4}

\usepackage{graphicx}% Include figure files

\begin{document}
\title{\Large \bf Nonuniversal $Z^{\prime}$ couplings in $B$ decays  }
\date{\today}
\author{\large \bf
Chuan-Hung~Chen$^{1,2}$\footnote{Email: physchen@mail.ncku.edu.tw}
and Hisaki Hatanaka$^{3}$\footnote{Email:hatanaka@phys.nthu.edu.tw}
} \affiliation{
$^{1}$Department of Physics, National Cheng-Kung University, Tainan, 701 Taiwan \\
$^{2}$National Center for Theoretical Sciences, Taiwan \\
$^{3}$Department of Physics, National Tsing-Hua University, Hsin-Chu
, 300 Taiwan }

\begin{abstract}
We study the impacts of nonuniversal $Z^{\prime}$ model, providing
flavor changing neutral current at tree level, on the branching
ratios (BRs), CP asymmetries (CPAs) and polarization fractions of
$B$ decays. We find that for satisfying the current data, the new
left- and right-handed couplings have to be included at the same
time. The new introduced effective interactions not only could
effectively explain the puzzle of small longitudinal polarization in
$B\to K^* \phi$ decays, but also provide a solution to the small CPA
of $B^{\pm}\to \pi^{0} K^{\pm}$. We also find that the favorable CPA
of $B^{\pm}\to \pi^{0} K^{\pm}$ is opposite in sign to the standard
model; meanwhile, the CPA of $B_{d}\to \pi^{0}K$ has to be smaller
than $-10\%$. In addition, by using the values of parameters which
are constrained by $B\to \pi K$, we find that the favorable ranges
of BRs, CPAs, longitudinal polarizations, and perpendicular
transverse polarizations for $(B^{\pm}\to \rho^{\pm} K^{*},\,
B_{d}\to \rho^{\mp} K^{*\pm})$ are $(17.1\pm 3.9,\,10.0\pm2.0)\times
10^{-6}$, $(3\pm 5,\, 21\pm 7)\%$, $(0.66\pm 0.10,\, 0.44\pm 0.08)$
and $(0.14\pm 0.10,\, 0.25\pm 0.09)$, respectively.

\end{abstract}
\maketitle

\section{Introduction}

Some puzzles have been found in $B$ meson decaying processes, such
as (a) the large branching ratios (BRs) of $B\to K \eta^{\prime}$
\cite{PDG04,th_etak}, (b) the small longitudinal polarizations of
$B\to K^* \phi$ decays \cite{bphik_pol,th_pol_1,th_pol_2,CG_PRD71}
and (c) the unmatched CP asymmetries (CPAs) and BRs in $B\to \pi K$
decays \cite{bkpi_cleo,bkpi_belle,bkpi_babar,th_pik,KOY_PRD72}. The
interesting thing is that the mazy problems are all related to the
flavor changing neutral current (FCNC) $b\to s$ decays. The
resultants push us not only to consider more precise QCD effects but
also to speculate the existence of new physics.

It is known that the FCNC processes usually arise from the loop
corrections in the standard model (SM) like models. The preference
of loop corrections originates from the strict constraints of $K$
meson oscillation. However, the constraints of $K$ system are only
on the first two generations, the FCNC at tree level in third
generation has no significant limit yet. In especial, the direct
constraint from $B_s-\bar{B}_s$ mixing, dictated by $b\to s$
interactions, is a lower bound in experiment \cite{PDG04}. Hence, it
will be interesting to investigate the models which FCNC occurs at
tree level and they could provide the solutions to the puzzles in
$b\to s$ decays.

One of simple extensions of SM for the effects of FCNC is the
unconventional $Z^{\prime}$ model, in which FCNC is arisen from the
family nonuniversal couplings \cite{Zpmodels}, i.e. the couplings of
$Z^{\prime}$ to different families are not the same. One of possible
ways to get the family nonuniversal couplings is to include an
additional $U(1)^{\prime}$ gauge symmetry; and then by the
requirement of anomaly free, the gauge charges for different
families are different \cite{DKW_PRD72}. Other models giving the
family nonuniversal $Z^{\prime}$ interactions could be referred to
Ref.~\cite{Zpmodels}. The detailed phenomenological analyses for
various low energy physics could be found in Ref. \cite{LP_PRD62}.
Especially, the implications on time-dependent CPA of $B\to K_{s}
\phi$ and on the BRs of $B\to\eta^{\prime} K$ have been studied by
Ref.~\cite{BCLL_PLB580}. Moreover, the solution to the $B\to \pi K$
puzzle by the nonuniversal $Z^{\prime}$ couplings is also discussed
by the authors of Ref.~\cite{BCLL_PLB598}. To further pursue the
effects on $B$ decaying processes, in this paper, we will take all
the measurements of $B\to K^{(*)} \phi$ and $B\to \pi K$ into
account to constrain the free parameters. By the constrained
parameters, we investigate the implications of the $Z^{\prime}$
model on BRs, CPAs and polarization fractions (PFs) for the decays
$B\to K^{*} \phi$, $B\to \pi K$ and $B\to \rho K^*$.

 For two-body color-allowed processes of $B$ decays, it is known that the dominant
hadronic effects are the factorized parts which could be simply
expressed as the multiplication of effective Wilson coefficients
(WCs), the decay constant, and the transition form factors, {\it
e.g.} $A \propto C(\mu) f_{M_1} F^{B\to M_{2}}(q^2=m^{2}_{M_1})$.
%%%%%%%%%%%%%%
Furthermore, by the observed CPA of O(10\%) in $B_{d}\to \pi^{\mp}
K^{\pm}$, we know that the large strong phases have to be
introduced. In order to self-consistently calculate the hadronic
effects, we employ the perturbative QCD (PQCD) approach
\cite{LP,KpPQCD} to evaluate the hadronic matrix elements, where the
large strong phases could be generated by the annihilation effects
of the effective operator $(V-A)\otimes (V+A)$.

It is known that in $B$ system which is composed of a heavy quark
and a light quark, the residual momentum of the light quark is
typically $k \sim O(m_{B}-m_{b})$ with $m_{B(b)}$ being the mass of
the B-meson (b-quark). In $B$ decays, if we regard that the
processes are dominated by short-distant interactions, for catching
up the energetic quark  from the b-quark decay to form a energetic
meson with the typical energy being $O(m_{B})$, the light quark
inside the $B$ meson has to obtain a large momentum from the b-quark
via the gluon exchange. Hence, the momentum transfer carried by the
hard gluon could be estimated to be $k_1-k_2$, where $k_1$ and
$k_{2}$ denote the momenta of spectator quarks inside the $B$ meson
and produced meson, respectively. In terms of light-cone
coordinates, the large components of $k_i\ (i=1,2)$
%and $k_2$
could be defined by $k^{+}_{i}=x_{i}
m_{B}/\sqrt{2}$
% and $k^{-}_{2}=x_{2} m_{B}/\sqrt{2} $
with $x_{i}$
%and $x_{2}$
being the momentum fractions. Hence, the squared
momentum of the exchanged hard gluon is $q^2=x_{1} x_{2} m^{2}_{B}$.
As known that the residual momentum of light quark in the $B$ meson is
$O(m_{B}-m_{b})$, $x_{1}$ is roughly
$O(m_{B}-m_{b})/m_{B}$. Since the produced light meson is energetic,
the momenta of valence quarks should be $O(m_{B}/\sqrt{2})$, i.e.
$x_{2}\sim O(1)$. By taking $x_{1}=0.16$, $x_{2}=0.5$ and
$m_{B(b)}=5.28(4.4)$ GeV, we get $\sqrt{q^2}\sim 1.5$ GeV. Since
the value reflects the typical reacting scale of $B$ decays in the
framework of the PQCD,  for the SM contributions, in our calculations
the values of weak WCs are estimated at the scale $\mu\sim 1.5$ GeV.%
%%%%%%%%%%%%%

The paper is organized as follows: In Sec.~\ref{sec:fcnc}, we
introduce the nonuniversal $Z^{\prime}$ effects for $b\to s$
transition. In Sec.~\ref{amplitudes}, based on the flavor diagrams,
we explicitly write out the factorizable amplitudes associated with
the new physics for the decays $B_d\to K^{(*)} \phi$, $B\to \pi K$
and $B\to \rho K^*$. In addition, we also define direct CPA and PFs.
Then by setting the values of parameters, in Sec.~\ref{sec:NA} we
give the calculated values for hadronic effects, present various
current experimental data for constraining the unknown parameters,
display the SM predictions and discuss the results of the
$Z^{\prime}$ model. Finally, we give a summary.

\section{FCNC for $b\to s$ transition in the $Z^{\prime}$ model
}\label{sec:fcnc}

In this section, we will introduce the neutral current interactions
in the SM and its extension with an extra $Z^{\prime}$ boson. Since
we will study the nonleptonic decays, in following discussions we
only concentrate on the quark sector. Although we concentrate on the
study of new physics, the used notation for new interacting
operators will be similar to those presented in the SM. Therefore,
it is useful to introduce the effective operators of the SM. Thus,
we describe the effective Hamiltonian for $b\to s q\bar{q}$ decays
as \cite{BBL}
\begin{equation}
H_{{\rm eff}}={\frac{G_{F}}{\sqrt{2}}}\sum_{q=u,c}V_{q}\left[
C_{1}(\mu) O_{1}^{(q)}(\mu )+C_{2}(\mu )O_{2}^{(q)}(\mu
)+\sum_{i=3}^{10}C_{i}(\mu) O_{i}(\mu )\right] \;,
\label{eq:hamiltonian}
\end{equation}
where $V_{q}=V_{qs}^{*}V_{qb}$ are the Cabibbo-Kobayashi-Maskawa
(CKM) \cite{CKM} matrix elements and the operators $O_{1}$-$O_{10}$
are defined as
\begin{eqnarray}
&&O_{1}^{(q)}=(\bar{s}_{\alpha}q_{\beta})_{V-A}(\bar{q}_{\beta}b_{\alpha})_{V-A}\;,\;\;\;\;\;
\;\;\;O_{2}^{(q)}=(\bar{s}_{\alpha}q_{\alpha})_{V-A}(\bar{q}_{\beta}b_{\beta})_{V-A}\;,
\nonumber \\
&&O_{3}=(\bar{s}_{\alpha}b_{\alpha})_{V-A}\sum_{q}(\bar{q}_{\beta}q_{\beta})_{V-A}\;,\;\;\;
\;O_{4}=(\bar{s}_{\alpha}b_{\beta})_{V-A}\sum_{q}(\bar{q}_{\beta}q_{\alpha})_{V-A}\;,
\nonumber \\
&&O_{5}=(\bar{s}_{\alpha}b_{\alpha})_{V-A}\sum_{q}(\bar{q}_{\beta}q_{\beta})_{V+A}\;,\;\;\;
\;O_{6}=(\bar{s}_{\alpha}b_{\beta})_{V-A}\sum_{q}(\bar{q}_{\beta}q_{\alpha})_{V+A}\;,
\nonumber \\
&&O_{7}=\frac{3}{2}(\bar{s}_{\alpha}b_{\alpha})_{V-A}\sum_{q}e_{q} (\bar{q}%
_{\beta}q_{\beta})_{V+A}\;,\;\;O_{8}=\frac{3}{2}(\bar{s}_{\alpha}b_{\beta})_{V-A}
\sum_{q}e_{q}(\bar{q}_{\beta}q_{\alpha})_{V+A}\;,  \nonumber \\
&&O_{9}=\frac{3}{2}(\bar{s}_{\alpha}b_{\alpha})_{V-A}\sum_{q}e_{q} (\bar{q}%
_{\beta}q_{\beta})_{V-A}\;,\;\;O_{10}=\frac{3}{2}(\bar{s}_{\alpha}b_{\beta})_{V-A}
\sum_{q}e_{q}(\bar{q}_{\beta}q_{\alpha})_{V-A}\;, \label{eq:ops}
\end{eqnarray}
with $\alpha$ and $\beta$ being the color indices. In Eq.
(\ref{eq:hamiltonian}), $O_{1}$-$O_{2}$ are from the tree level of
weak interactions, $O_{3}$-$O_{6}$ are the so-called gluon penguin
operators and $O_{7}$-$O_{10}$ are the electroweak penguin
operators, while $C_{1}$-$C_{10}$ are the corresponding WCs. Using
the unitarity condition, the CKM matrix elements for the penguin
operators $O_{3}$-$O_{10}$ can also be expressed as
$V_{u}+V_{c}=-V_{t}$.

 For studying the $Z^{\prime}$ model, as usual we describe the Lagrangian for the neutral current
interactions in terms of weak eigenstates as
\cite{LP_PRD62,LL_PRD45}
\begin{eqnarray}
{\cal L}_{NC}&=& - g_{1} J^{\mu}_{1} Z^{0}_{1\mu} -g_{2} J^{\mu}_{2}
Z^{0}_{2\mu}, \nonumber \\
     J^{\mu}_{1}&=& \bar{q}_{i} \gamma^{\mu} [ \epsilon^{L}_{q} P_{L} +
\epsilon^{R}_{q} P_{R}] q_{i}, \nonumber \\
     J^{\mu}_{2}&=& \bar{q}_{i} \gamma^{\mu} [ (\tilde{\epsilon}^{L}_{qL})_{ij} P_{L} +
(\tilde{\epsilon}^{R}_{qR})_{ij} P_{R}] q_{i}.
\label{eq:3}
\end{eqnarray}
where the subscript of $q_{i}$ denotes flavor index of quark,
% the chiral projection operators
 $P_{L,R}=(1\mp\gamma_{5})/2$,
%In Eq. (\ref{eq:3}),
$Z^{0}_{1}$ and $Z^{0}_{2}$ are the neutral gauge bosons,
corresponding to $SU(2)\times U(1)$ and the
extended Abelian gauge symmetries,
 and $g_{1}$ and $g_{2}$ are the associated gauge couplings, respectively.
 %; clearly, the former corresponds to the neutral gauge boson in
%$SU(2)\times U(1)$ while the
%latter stands for the neutral gauge boson associated with the
%extended Abelian gauge symmetry.
We note that $\epsilon^{L,R}_q$ are universal couplings of the SM
while $\tilde{\epsilon}^{L,R}_q$ are $3\times 3$ matrices and denote
the effects of nonuniversal couplings. In general, the $Z^{0}_{1}$
and $Z^{0}_{2}$ bosons will mix each other so that the physical
states of $Z$ bosons could be parametrized by
\begin{eqnarray}
\left(
  \begin{array}{c}
    Z \\
    Z^{\prime}\\
  \end{array}
\right)= \left(
           \begin{array}{cc}
             \cos\theta & \sin\theta \\
             -\sin\theta & \cos\theta  \\
           \end{array}
         \right) \left(
                   \begin{array}{c}
                     Z^{0}_{1} \\
                     Z^{0}_{2} \\
                   \end{array}
                 \right)
\end{eqnarray}
where $\theta$ denotes the $Z-Z^{\prime}$ mixing angle. In addition,
the physical states of quarks could be related to the weak
eigenstates by $U^{p}_{L(R)}=V_{UL(R)}U^{w}$ and
$D^{p}_{L(R)}=V_{DL(R)}D^{w}$ in which $U^{T}=(u,c,t)$,
$D^{T}=(d,s,b)$, and the $V_{UL(R)}$ and $V_{DL(R)}$ are the unitary
matrices for diagonalizing weak states to physical eigenstates. The
CKM matrix is defined by $V_{CKM}=V_{UL} V^{\dagger}_{DL}$. As a
result, in terms of physical states the effective interactions for
$b \to s$ decays could be written as
\begin{eqnarray}
{\cal H}_{Z}&=&\frac{8G_F}{\sqrt{2}} \sin\theta\cos\theta \left(
\frac{g_2}{g_1}\right) \sum_{\chi_1, \chi_2}\sum_{q}\left[
B_{sb}^{\chi_2 *} \epsilon_{q}^{\chi_1} \bar{b} \gamma_{\mu}
P_{\chi_2} s \, \bar{q} \gamma^{\mu} P_{\chi_1}
q \right] + h.c. \; ,\nonumber \\
  %%%%%%%%%%%%%%%%%%%
      {\cal H}_{Z^{\prime}}&=&\frac{8G_F}{\sqrt{2}} \cos^2\theta \left(
\frac{g_2 m_{Z}}{g_1 m_{Z^{\prime}}}\right)^2
\sum_{\chi_1,\chi_2}\sum_{q}\left[ B_{sb}^{\chi_2 *} B_{qq}^{\chi_1}
\bar{b} \gamma_{\mu} P_{\chi_2} s \, \bar{q} \gamma^{\mu} P_{\chi_1}
q \right] + h.c. \label{zp}
\end{eqnarray}
where ${\cal H}_{Z}$ and ${\cal H}_{Z^{\prime}}$ express the effects
of $Z-Z^{\prime}$ mixing and $Z^{\prime}$, respectively, the $q$
could be $u,\ d,\ s$ and $c$ quark, $\chi_{i=1,\, 2}=L$ and $R$, and
\begin{eqnarray}
  \epsilon_q^{L}&=&T^{q}_{3}-Q_q\sin^2\theta_W ,\ \ \
  \epsilon_{q}^{R}=- Q_q \sin^2\theta_W, \nonumber \\
  B^{\chi}_{DD}&=&V^{D}_{\chi} \tilde{\epsilon}^{\chi}_{D} V^{D\dagger}_{\chi},
  \ \ \ B^{\chi}_{UU}=V^{U}_{\chi} \tilde{\epsilon}^{\chi}_{U}
  V^{U\dagger}_{\chi}.
\end{eqnarray}
Here, the capital $UU$ and $DD$ in the subscript of $B^{\chi}$
parameter could be the flavors (u, c) and (d, s, b), respectively,
and the $\theta_{W}$ is the Weinberg's angle. By current
experimental data, it is known that the mixing angle $\theta$ is
limited to be less than $O(10^{-3})$ \cite{BCLL_PLB580,BCLL_PLB598}.
If the mass of $Z^{\prime}$ is in the range of a few hundred GeV to
1 TeV, the dominant effects only come from the $Z^{\prime}$
exchange. Therefore, under this assumption, we will neglecte the
contributions of $Z-Z^{\prime}$ mixing.

According to the interactions of Eq. (\ref{zp}), the new effective
Hamiltonian for $b\to s q \bar{q}$ decays could be written as
   \begin{eqnarray}
      H^{Z^{\prime}}_{\rm eff}&=&-\frac{G_F}{\sqrt{2}}V^*_{tb} V_{ts} \sum_q \left\{
      (\bar{b} s)_{V-A} \left[ \left(\Delta C_3 + \Delta C_9 \frac{3}{2}e_{q} \right) (\bar{q} q)_{V-A}
      \right. \right.\nonumber \\
      && \left. +\left( \Delta C_5 + \Delta C_7 \frac{3}{2} e_q \right) (\bar{q} q)_{V+A}
      \right] +
         (\bar{b} s)_{V+A} \left[ \left(\Delta C^{\prime}_3
         + \Delta C^{\prime}_9 \frac{3}{2}e_{q} \right) (\bar{q}
         q)_{V+A} \right. \nonumber \\
      && + \left. \left. \left( \Delta C^{\prime}_5 + \Delta C^{\prime}_7 \frac{3}{2} e_q \right) (\bar{q} q)_{V-A}
      \right]      \right\} \label{effint}
   \end{eqnarray}
with
   \begin{eqnarray}
    \Delta C_{3[5]}&=&-\frac{2}{3}\left( \frac{g_2 M_Z}{g_1
    M_{Z^{\prime}}}\right)^2 \frac{1}{V^*_{tb}V_{ts}} B^{L*}_{sb} \left( B^{L[R]}_{UU}+2
    B^{L[R]}_{DD}\right), \nonumber \\
       \Delta C^{\prime}_{3[5]}&=&-\frac{2}{3}\left( \frac{g_2 M_Z}{g_1
    M_{Z^{\prime}}}\right)^2 \frac{1}{V^*_{tb}V_{ts}} B^{R*}_{sb} \left( B^{R[L]}_{UU}+2
    B^{R[L]}_{DD}\right), \nonumber \\
                   \Delta C_{9[7]}&=&-\frac{4}{3}\left( \frac{g_2 M_Z}{g_1
    M_{Z^{\prime}}}\right)^2 \frac{1}{V^*_{tb}V_{ts}} B^{L*}_{sb} \left(
    B^{L[R]}_{UU}-
    B^{L[R]}_{DD}\right),\nonumber \\
                              \Delta C^{\prime}_{9[7]}&=&-\frac{4}{3}\left( \frac{g_2 M_Z}{g_1
    M_{Z^{\prime}}}\right)^2 \frac{1}{V^*_{tb}V_{ts}} B^{R*}_{sb} \left(
    B^{R[L]}_{UU}-B^{R[L]}_{DD}\right). \label{wcs}
           \end{eqnarray}
The expressions have been written as the four-fermion operators of
the SM, shown in Eq. (\ref{eq:ops}). The operators associated with
the new unprimed WCs $\Delta C_{3,5,7,9}$ are the same as SM.
However, the operators associated with primed coefficients $\Delta
C^{\prime}_{3,5,7,9}$ have different chirality from those in the SM
for $b-s$ couplings. That is, the flavor-changing (FC) $Z^{\prime}$
model provides different chiral flavor structures for FCNC
processes. It has been found that the new effective WCs of Eq.
(\ref{wcs}) could be simplified if one assumes $B^{\chi}_{UU} \simeq
-2 B^{\chi}_{DD}$ \cite{BCLL_PLB598}. Although in general the
assumption is unnecessary, for simplicity, we still impose the
condition in our case. Hence, we get $\Delta C^{(\prime)}_{3[5]}
\approx 0$ and
  \begin{eqnarray}
         \Delta C_{9[7]}&=&4\left( \frac{g_2 M_Z}{g_1
    M_{Z^{\prime}}}\right)^2 \frac{1}{V^*_{tb}V_{ts}} B^{L*}_{sb}
    B^{L[R]}_{DD},\nonumber \\
            \Delta C^{\prime}_{9[7]}&=&4\left( \frac{g_2 M_Z}{g_1
    M_{Z^{\prime}}}\right)^2 \frac{1}{V^*_{tb}V_{ts}} B^{R*}_{sb}
    B^{R[L]}_{DD}.\label{eq:c97}
  \end{eqnarray}

Although the hadronic matrix elements, describing the $B$ decaying
to two final mesons through the effective Hamiltonian, depend on the
chiral and color structures of four-fermion operators, we find that
the associated effective WCs could be classified and reexpressed to
be more useful form by
\begin{eqnarray}
  a_{1} &=& C_{2}+\frac{C_1}{N_c}, \ \ \ a_{2}= C_1+
      \frac{C_2}{N_c}, \nonumber \\
     a^{q}_3&=&C_{3}+\frac{C_{4}}{N_c}+ \frac{3}{2} e_{q} \left(C^{NP}_9
     +\frac{C_{10}}{N_c} \right), \ \ \ a^{q}_4 =C_{4}+\frac{C_{3}}{N_c}+
     \frac{3}{2} e_{q} \left(C_{10} +\frac{C^{NP}_{9}}{N_c} \right),\nonumber \\
a^{q}_5&=&C_{5}+\frac{C_{6}}{N_c}+ \frac{3}{2} e_{q}
          \left(C^{NP}_7+\frac{C_{8}}{N_c} \right),\ \ \
                  a^{q}_6=C_{6}+\frac{C_{5}}{N_c}+ \frac{3}{2} e_{q}
                  \left(C_8+\frac{C^{NP}_{7}}{N_c} \right), \nonumber \\
             a^{\prime qNP}_3 &=& \frac{3}{2} e_{q} \Delta C^{\prime}_9, \ \ \
             a^{\prime qNP}_4= \frac{3}{2} e_{q} \frac{\Delta
             C^{\prime}_9}{N_c},\ \ \
             a^{\prime qNP}_5 = \frac{3}{2} e_{q} \Delta C^{\prime}_7, \ \ \
              a^{\prime qNP}_6= \frac{3}{2} e_{q} \frac{\Delta
             C^{\prime}_7}{N_c} \label{effwcs}
\end{eqnarray}
with $C^{NP}_{9(7)}=C_{9(7)}+\Delta C_{9(7)}$. The superscript $q$
of Eq. (\ref{effwcs}) denotes the corresponding flavor and $e_{q}$
is its charge. Since the considering $Z^{\prime}$ model has the
flavor structures which are the same as SM, to be more clear to
understand the influence of new physics, we rewrite Eq.
(\ref{effwcs}) to be
  \begin{eqnarray}
a^{q}_3&=& a^{qSM}_{3}+ \frac{3}{2} e_{q} \Delta C_{9} , \ \ \
      a^{q}_4=a^{qSM}_{4}+\frac{3}{2} e_{q} \frac{\Delta C_{9}}{N_c}
,\nonumber \\
          a^{q}_5&=& a^{qSM}_{5}+ \frac{3}{2} e_{q} \Delta C_7,\ \ \
                            a^{q}_6 = a^{qSM}_{6} + \frac{3}{2} e_{q} \frac{\Delta
                  C_7}{N_c}. \label{aeff}
             \end{eqnarray}

\section{Decay amplitudes for $B\to \phi K^{(*)}$ and  $B\to \pi(\rho) K^{(*)}$
decays}\label{amplitudes}

\begin{figure}[htbp]
\includegraphics*[width=3.in]{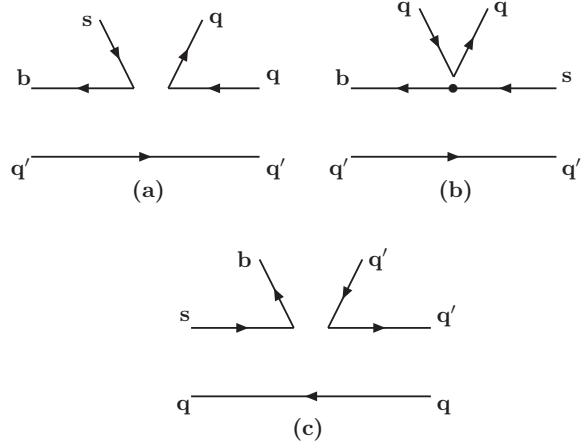}  \caption{The flavor diagrams for $b\to s q \bar{q}$: (a) and (b) stand for
the emission topologies while (c) is annihilation topology.  }
 \label{fig:flavor}
\end{figure}

To describe the amplitudes for $B$ decays, we have to know  not only
the relevant effective weak interactions but also all possible
topologies for the specific process. We display the general
involving flavor diagrams for $b\to s q \bar{q}$ in Fig.
\ref{fig:flavor}, where (a) and (b) denote the emission topologies
while (c) is the annihilation topology. The flavor $q$ in
Fig.\ref{fig:flavor}(a) and (b) is produced by gauge bosons and
could be $u$, or $d$ or $s$ quark if the final states are the light
mesons; however, $q^{\prime}$ stands for the spectator quark and
could only be $u$ or $d$ quark, depending on the $B$ meson being
charged or neutral one.
%%%%%%%%%%%%%%%%%%%%%%%%%%%%%%%%%%%%%%%%%%%%%%%%%%%%%%%%%%%%%%%%%%%%%%%%%
%%%%%%%%%%%%%%%%%%%%%%%%%%%%%%%%%%%%%%%%%%%%%%%%%%%%%%%%%%%%%%%%%%%%%%%%%
However, the role of $q$ and $q^{\prime}$ in
Fig.~\ref{fig:flavor}(c) is reversed so that $q=u$, or $d$ or $s$ is
the spectator quark and $q^{\prime}=u$ or $d$ is dictated by gauge
interactions. We note that the presented flavor diagrams are based
on the penguin operators of the SM. Except the different type of
interactions at vertices, the flavor diagrams induced by new physics
should be similar to those generated by the SM. In addition, since
the matrix elements obtained by the Fierz transformation of
$O_{3,4}$ are the same as those of $O_{1,2}$, we don't further
consider the matrix elements of tree operators. Hence, in terms of
the effective interactions of the SM and those shown in Eq.
(\ref{effint}), the expressions of Eqs. (\ref{effwcs}) and
(\ref{aeff}), and the flavor diagrams of Fig.~\ref{fig:flavor}, we
can investigate the decay amplitudes for $B\to K \phi$, $B\to \pi
K$, $B\to K^* \phi$ and $B\to \rho K^*$.

\subsection{$B_{d}\to K^{0} \phi$}\label{sec:bkphi}

Although there are  charged and neutral modes in $B\to K \phi$
decays, since the differences of flavor diagrams in charged and
neutral modes are only the parts of small tree annihilation, we only
concentrate on the decay $B_{d}\to K^{0} \phi$. At quark level, the
process is controlled by the decay $b\to s s \bar{s}$, therefore,
$q=s$ and $q^{\prime}=d$ in Fig.~\ref{fig:flavor}(a) and (b), but
they are reversed in Fig.~\ref{fig:flavor}(c). Hence, the decay
amplitude for $B_d\to K^0 \phi$ is written as
\begin{eqnarray}
{\cal M}^{Z^{\prime}}_{K\phi }&=&V^{*}_{tb}V_{ts}\left[ f_{\phi}
\xi_{1} F^{e}_{1K} + f_{B} (\xi_{2} F^{a}_{1 K\phi}+  \xi_{3}
F^{a}_{2 K\phi})\right]
 \end{eqnarray}
and
\begin{eqnarray}
\xi_{1}&=& a^s +  a^{\prime s NP},\ \ \
          \xi_{2[3]}= a^d_{4[6]} +
          a^{\prime d NP}_{4[6]}, \label{eq:phik_xi}
              %                  \xi_{3L}=\xi_{3\parallel}&=& a^d_{6} + a^{dNP}_6 -
%          a^{\prime d NP}_6, \nonumber \\
%                   \xi_{3\perp}&=& a^d_{6} + a^{dNP}_6 +
%          a^{\prime d NP}_6,
\end{eqnarray}
with $a^{s}=a^{s}_{3}+a^{s}_{4}+a^{s}_5$ and $a^{\prime
sNP}=a^{\prime sNP}_{3}+a^{\prime sNP}_{4}+a^{\prime sNP}_5$. The
$f_{\phi}$ means the decay constant of $\phi$ meson and is defined
by $\langle 0 | \bar{s} \gamma_{\mu}s | \phi\rangle=f_{\phi}
m_{\phi} \epsilon_{\mu}$, in which $m_{\phi}$ and $\epsilon_{\mu}$
express the mass and polarization vector of $\phi$ meson. The
hadronic matrix element $F^{e}_{1K}$ is from the diagrams (a) and
(b). However, the $F^{a}_{1K\phi}$ and $F^{a}_{2K\phi}$ come from
the annihilation topology diagram (c). The detailed expressions of
the hadronic matrix elements are given in the
Appendix~\ref{app:hardamp_pv}. The decay rate for $B\to P V$ is
written as
\begin{equation}
\Gamma =\frac{G_{F}^{2}m^{2}_{B} P_c}{16\pi } |{\cal M}|^2
\end{equation}
where $P_c\equiv m^2_{1} m^{2}_{2}(r^2-1)/m^2_{B}$ with
$r=P_{1}\cdot P_{2}/(m_{1}m_{2})$ is the momentum of the outgoing
vector mesons. By the decay width, we can also define the direct CPA
to be
\begin{eqnarray}
A_{CP}=\frac{\bar{\Gamma} -\Gamma}{\bar{\Gamma}+\Gamma}
\end{eqnarray}
where $\bar{\Gamma}$ is the decay rate of antiparticle.

\subsection{$B\to \pi K$} \label{sec:bpik}
There are four specific modes in $B\to \pi K$ decays. Since all BRs
and CPA of $B_d\to \pi^{-} K^{+}$ are observed well in experiments,
we have to analyze all of them in detail. We begin the analysis from
the decay $B^{+}\to \pi^{+} K^{0}$. According to the flavor diagrams
Fig.~\ref{fig:flavor}(a) and (c), the emission and annihilation
topologies of the decay are described by $q=d$ and $q^{\prime}=u$.
Hence, the decay amplitude for $B^{+} \to \pi^{+} K^{0} $ decay
could be expressed by
\begin{eqnarray}
{\cal M}^{Z^{\prime}}_{\pi^+K^0}&=& V^{*}_{tb}V_{ts}\left[f_{K}
(\zeta^{d}_{1} F^{e}_{1\pi } + \zeta^{d}_{2} r_{K} F^{e}_{2\pi }) +
f_{B} (\zeta^{u}_{1} F^{a}_{1\pi K}+
 \zeta^{u}_{2} F^{a}_{2\pi K})\right],
\end{eqnarray}
where $f_{K}$ is the decay constant of kaon and defined by $\langle
0| \bar{s} \gamma_{\mu} \gamma_{5} d |K^{0}\rangle =f_{K} p_{\mu}$,
$r_{K}=m^{0}_{K}/m_{B}$ with $m^{0}_{K}$ being associated with
$\langle 0| \bar{s} \gamma_{5} d |K^{0}\rangle =f_{K} m^{0}_{K}$,
and
  \begin{eqnarray}
     \zeta^{q}_{1}&=&a^{q}_{4}-a^{\prime qNP}_{4} , \ \ \
     \zeta^{q}_{2}=a^{q}_{6}-a^{\prime qNP}_{6}. \label{eq:zeta}
  \end{eqnarray}
The hadronic matrix elements $F^{e}_{1\pi}$, $F^{a}_{1\pi K }$ and
$F^{a}_{2\pi K}$ are similar to those for $B_d\to K^0 \phi$ and the
detailed expressions are given in Appendix~\ref{app:hardamp_pp}. In
addition, we have a new contribution $F^{e}_{2\pi K}$ which arises
from the emission topologies of $O_{6,8}$.

Similar to $B^{+}\to \pi^{+} K^0$, we can obtain the decay amplitude
of $B_{d}\to \pi^{-} K^+$ easily by using $q(q^{\prime})=u(d)$
instead of $q(q^{\prime})=d(u)$. Thus, the decay amplitude for
$B_d\to \pi^{-} K^+$ decay is written as
\begin{eqnarray}
{\cal M}^{Z^{\prime}}_{\pi^+K^-}&=& V^{*}_{tb}V_{ts}\left[f_{K}
(\zeta^{u}_{1} F^{e}_{1\pi} + \zeta^{u}_{2} r_{K} F^{e}_{2\pi}) +
f_{B} (\zeta^{d}_{1} F^{a}_{1\pi K}+
 \zeta^{d}_{2} F^{a}_{2\pi K})\right] \nonumber \\
 &&- V^*_{ub} V_{us} f_{K} a_{1}F^{e}_{1\pi },
\end{eqnarray}
where we have included the tree contributions. As mentioned before,
except the CKM matrix elements and effective WCs, the hadronic
effects of tree are the same as the penguin operators $O_{3,4}$.
Hence, the hadronic effects encountered in $B_{d}\to \pi^{-} K^{+}$
decay are the same as those in $B^{+}\to \pi^{+} K^0$.

Next, we analyze the situation of $B_{d}\to\pi^{0} K^0$ decay. one
can easily find that besides the involving flavor diagrams appeared
in the decays $B^{+}\to \pi^{+} K^0$ and $B_{d}\to \pi^{-} K^+$, the
diagram Fig.~\ref{fig:flavor}(b) corresponding to the electroweak
penguin contributions of the SM should be also included. Taking
proper flavors for $q$ and $q^{\prime}$, the decay amplitude for
$B_d\to \pi^{0} K^0$ decay is given by
\begin{eqnarray}
\sqrt{2}{\cal M}^{Z^{\prime}}_{\pi^0K^0}&=&
V^{*}_{tb}V_{ts}\left[-f_{K} (\zeta^{d}_{1} F^{e}_{1\pi } +
\zeta^{d}_{2} r_{K} F^{e}_{2\pi }) + f_{\pi} \zeta F^{e}_{1K}- f_{B}
(\zeta^{d}_{1} F^{a}_{1\pi K}+
 \zeta^{d}_{2} F^{a}_{2\pi K})\right] \nonumber \\
 && - V^*_{ub} V_{us} f_{\pi} a_{2} F^{e}_{1 K},
\end{eqnarray}
with $\zeta=a^{u}_{3}-a^{d}_{3}+a^{u}_{5}-a^{d}_{5} -(a^{\prime
uNP}_{3}-a^{\prime dNP}_{3}+a^{\prime uNP}_{5}-a^{\prime dNP}_{5})$.
We note that the new term $f_{\pi} F^{e}_{1K}$, corresponding to
Fig.~\ref{fig:flavor}(b), has opposite in sign to other terms. The
reason comes from the flavor wave function of $\pi^{0}$ being
$(\bar{u} u-\bar{d} d)/\sqrt{2}$. The Fig.~\ref{fig:flavor}(b) picks
both components while others only take $\bar{d} d$ component. Since
the tree contributions are color suppressed, the corresponding WC is
$a_{2}$.

After introducing the decay amplitudes of $B^{+}\to \pi^{+} K^0$,
$B_d\to \pi^{-} K^+$ and $B_d\to \pi^0 K^0 $, the amplitude for
$B^+\to \pi^{0} K^+$ decay could be immediately obtained as
\begin{eqnarray}
\sqrt{2}{\cal
M}^{Z^{\prime}}_{\pi^0K^+}&=&V^{*}_{tb}V_{ts}\left[f_{K}
(\zeta^{u}_{1} F^{e}_{1\pi } + \zeta^{u}_{2} r_{K} F^{e}_{2\pi }) +
f_{\pi} \zeta F^{e}_{1 K}+ f_{B} (\zeta^{u}_{1} F^{a}_{1\pi K}+
 \zeta^{u}_{2} F^{a}_{2\pi K}) \right]\nonumber \\
 && - V^*_{ub} V_{us}  (f_{K} a_1 F^{e}_{1\pi}+ f_{\pi} a_{2} F^{e}_{1
 K}).
\end{eqnarray}
Clearly, the amplitudes shown in the first three decay modes all
appear in the decay $B^+\to \pi^0 K^+$. That is, once one determines
the first three decays, the decay $B^{+}\to \pi^0 K^+$ is also
fixed.

\subsection{$B_{d}\to K^{*0} \phi$}

For the production of two vector mesons in $B$ decays, since both
vector mesons carry spin degrees of freedom, the decay amplitudes
are related to not only the longitudinal parts but also transverse
parts. In terms of the notation of Ref.~\cite{CKLPRD66}, the
amplitude ${\cal M}^{(h)}$ could be expressed by
\begin{eqnarray}
{\cal M}^{(h)}
% \epsilon_{1\mu}^{*}(h)\epsilon_{2\nu}^{*}(h)
%\left[ a \,g^{\mu\nu} + {b \over M_{V_1} M_{V_2}} P_2^\mu P_1^\nu +
%i{c \over M_{V_1} M_{V_2} } \epsilon^{\mu\nu\alpha\beta} P_{1\alpha}
%P_{2\beta}\right]\;
%\nonumber \\
\equiv m_{B}^{2}{\cal M}_{L} + m_{B}^{2}{\cal M}_{N}
\epsilon^{*}_{1}(t)\cdot\epsilon^{*}_{2}(t) +i{\cal
M}_{T}\epsilon^{\alpha \beta\gamma \rho}
\epsilon^{*}_{1\alpha}(t)\epsilon^{*}_{2\beta}(t) P_{1\gamma
}P_{2\rho } \label{eq:hamp}
\end{eqnarray}
with the convention $\epsilon^{0123} = 1$, where the superscript $h$
is the helicity, the subscript $L$ stands for $h=0$ component while
$N$ and $T$ express another two $h=\pm 1$ components, and
$\epsilon^*_{1}(t)\cdot \epsilon^{*}_{2}(t)=1$ with $t=\pm 1$.
Hence, each helicity amplitude could be written as \cite{CKLPRD66}
\begin{eqnarray}
H_{0}&=&m^{2}_{B} {\cal M}_L\;,
\nonumber\\
H_{\pm}&=& m^{2}_{B} {\cal M}_{N} \mp  m_{V_1} m_{V_2}
\sqrt{r^2-1}{\cal M}_{T}\;.
\end{eqnarray}
In addition, we can also write the amplitudes in terms of
polarizations as
\begin{eqnarray}
A_{L}=H_{0} \ \ \ A_{\parallel(\perp)}=\frac{1}{\sqrt{2}}(H_{-} \pm
H_{+}). \label{pol-amp}
\end{eqnarray}
Accordingly, the PFs can be defined as %
 \begin{eqnarray}
R_i=\frac{|A_i|^2}{|A_L|^2+|A_{\parallel}|^2+|A_{\perp}^2|}\,,\ \
(i=L,\parallel,\perp)\,, \label{eq:pol}
 \end{eqnarray} %
Consequently, the decay rate for $B\to V_{2} V_{1}$ is given by
\begin{equation}
\Gamma =\frac{G_{F}^{2}P_c}{16\pi m^{2}_{B} } \left[ |A_{L}|
^{2}+|A_{\parallel}|^{2} + | A_{\perp}|^{2}\right]\; \label{dr1}
\end{equation}
where $P_c\equiv |P_{1z}|=|P_{2z}|$ is the momentum of either of the
outgoing vector mesons.

 From Fig.~\ref{fig:flavor}, it is easy to see that the associated
flavor diagrams for $B\to K^* \phi$ are the same as those for $B\to
K \phi$ decays. Furthermore, as the results for the neutral and
charged modes are expected to be similar by neglecting the small
annihilation  contributions from tree operators $O^{u}_{1,2}$
appearing in the charged mode, which will be discussed in
Sec.~\ref{sec:tinput}.
%That is, except the contributions from tree operators $O^{u}_{1,2}$
%appearing in the charged mode, the neutral and charged modes have
%similar results. However, as known that the effects of $O^{u}_{1,2}$
% belong annihilation contributions and negligible,
%see the discussions of Sec.~\ref{sec:tinput}.
%Therefore,
We will
%still
concentrate on the neutral $B$ decay. In terms of the distribution
amplitudes of vector mesons, defined in Appendix~\ref{app:distramp},
the decay amplitudes with various helicities defined by Eq.
(\ref{eq:hamp}) are given by
\begin{eqnarray}
{\cal M}^{Z^{\prime}}_{ K^*H}&=&V^{*}_{tb}V_{ts}\left[f_{\phi}
\xi_{1H} F^{e}_{ K^* H} + f_{B} \xi_{2H} F^{a}_{1 K^*\phi H}+ f_{B}
 \xi_{3H} F^{a}_{ 2K^*\phi H}\right]
\end{eqnarray}
where $H=L,\,N,\, T$ and
\begin{eqnarray}
\xi_{1L}=\xi_{1\parallel}&=& a^s  - a^{\prime s NP},\ \ \
\xi_{1\perp}= a^s + a^{\prime s NP}
\nonumber \\
            \xi_{2[3]L}=\xi_{2[3]\parallel}&=& a^d_{4[6]}  -
          a^{\prime d NP}_{4[6]}, \ \ \
              \xi_{2[3]\perp}= a^d_{4[6]} +
          a^{\prime d NP}_{4[6]}. \label{eq:phiks_xi}
%                  \xi_{3L}=\xi_{3\parallel}&=& a^d_{6} + a^{dNP}_6 -
%          a^{\prime d NP}_6, \nonumber \\
%                   \xi_{3\perp}&=& a^d_{6} + a^{dNP}_6 +
%          a^{\prime d NP}_6,
\end{eqnarray}
The definitions of $a^{s}$ and $a^{\prime sNP}$ are the same as
those for $B_{d}\to K^0 \phi$ decay. The explicit expressions for
$\{F^{e},F^{a}\}$ could be referred to
Appendix~\ref{app:hardamp_vv}.

\subsection{$B\to \rho K^*$} \label{sec:brhoks}

Since the quark compositions of $\rho$ and $K^*$ mesons are the same
as those of $\pi$ and $K$, respectively,  the flavor diagrams for
$B\to \rho K^*$ and $B\to \pi K$ decays should be the same. However,
due to $\langle V(p) | \bar{q} q^{\prime}| 0\rangle \propto
\epsilon_{V}(p) \cdot p=0$ in which the scalar vertex is arisen from
the Fierz transformation of $(V-A)\otimes (V+A)$, the emitted
factorizable contributions of four-fermion operators $O_{6,8}$ are
vanished, i.e. $a^q_{6}$ have no contributions. Consequently, it
could be expected that  BRs of $B\to \rho K^*$ are smaller than
those of $B\to \pi K$ in the SM.

Although there are four possible modes in the $B\to \rho K^*$
decays, we only concentrate on the decays $B^{+}\to \rho^{+} K^{*0}$
and $B^{0}\to \rho^{-} K^{*+}$ that they have larger BRs. Following
the definition of Eq. (\ref{eq:hamp}), the various helicity
amplitudes for $B^{+}\to \rho^{+} K^{*0}$ decay would be written as
       \begin{eqnarray}
{\cal M}^{Z^{\prime}}_{\rho^+
K^{*0}H}&=&V^{*}_{tb}V_{ts}\left[f_{K^*} \zeta^{d}_{1H} F^{e}_{\rho
 H} + f_{B} \zeta^{u}_{2H} F^{a}_{1\rho K^* H}+ f_{B}
 \zeta^{u}_{3H} F^{a}_{2\rho K^* H}\right].
\end{eqnarray}
The associated effective WCs are given by
\begin{eqnarray}
       \zeta^{q}_{1H=L,\parallel}&=&\zeta^{q}_{2H=L,\parallel}=\zeta^{q}_{1},
       \ \ \ \zeta^{q}_{3H=L,\parallel}= \zeta^{q}_{2}, \nonumber
       \\
       \zeta^q_{1\perp}&=& \zeta^q_{2\perp}=a^{q}_{4}+ a^{\prime qNP}_{4}, \ \ \
         \zeta^q_{3\perp}= a^{q}_{6}+ a^{\prime qNP}_{6}.
         \label{eq:zeta_rhoks}
\end{eqnarray}
Similarly, the decay amplitude for $B^{0}\to \rho^{-} K^{*+}$ decay
is written as
       \begin{eqnarray}
{\cal M}^{Z^{\prime}}_{\rho^-
K^{*+}H}&=&V^{*}_{tb}V_{ts}\left[f_{K^*} \zeta^{u}_{1H} F^{e}_{\rho
H} + f_{B} \zeta^{d}_{2H} F^{a}_{1\rho K^* H}+ f_{B}
 \zeta^{d}_{3H} F^{a}_{2\rho K^* H}\right]\nonumber \\
 && -V^{*}_{ub}V_{us} a_{1} f_{K^*}
 F^{e}_{\rho H}.
\end{eqnarray}
The definitions of $\zeta^{q}_{1(2)}$ are the same as those for
$B\to \pi K$, expressed by Eq. (\ref{eq:zeta}).

\section{Numerical Analysis}\label{sec:NA}

\subsection{Theoretical inputs}\label{sec:tinput}

To obtain numerical estimations, the values of theoretical
parameters in the SM related to the weak interactions are taken as
follows: $G_F=1.166\times 10^{-5}$ GeV$^{-2}$, $V_{us}=0.224$,
$V_{ts}=0.041$, $V_{ub}=3.5\times 10^{-3} e^{-i\phi_{3}}$ with
$\phi_{3}=72^{0}$. The decay constants of mesons are set to be
$f_{\pi}=130$, $f_{K}=160$, $f_{B}=190$, $f^{(T)}_{\phi}=237 (170)$,
and $f^{(T)}_{K^*}=f^{(T)}_{\rho}=200 (160)$ MeVs. The lifetimes of
charged and neutral $B$ mesons are chosen as $\tau_{B^{+}}=1.67
\times 10^{-12}\;s$ and $\tau_{B_d}=1.56\times 10^{-12}\; s$,
respectively. Since we use PQCD approach to calculate the hadronic
matrix elements, we set the scale of weak WCs at $\mu=
\sqrt{\bar{\Lambda} m_{B}}\approx 1.5$ GeV, therefore, the values of
SM shown in Eq. (\ref{effwcs}) are estimated to be
\begin{eqnarray}
a_1&=&1.07, \ \ a_2=-0.028, \nonumber \\
a^{uSM}_3&=& -0.002, \ \ a^{uSM}_4=-0.036, \ \ a^{uSM}_5=-0.008, \ \
a^{uSM}_6=-0.055, \nonumber \\
a^{dSM}_3&=& 0.012, \ \ a^{dSM}_4=-0.036, \ \ a^{dSM}_5=-0.008, \ \
a^{dSM}_6=-0.056,
\end{eqnarray}
and $a^{sSM}=a^{dSM}$. In addition, in Table~\ref{tab:hardfun} we
present the values of the hadronic effects, which are displayed in
Secs. \ref{sec:bkphi}$-$\ref{sec:brhoks} and calculated by PQCD
approach.
\begin{table}[hptb]
\caption{The values of factorizable amplitudes.} \label{tab:hardfun}
\begin{ruledtabular}
\begin{tabular}{ccccc}
$F^{e}_{1K}$  & $F^{a}_{1 K\phi}$ & $F^{a}_{2 K\phi}$& $F^{e}_{1\pi}$ & $F^{e}_{2\pi}$ \\
\hline
  0.37 & $(-9.88+i7.54)10^{-4}$ & $-0.047+i0.14$ & 0.24 & 0.50 \\ \hline \hline
$F^{a}_{1\pi K}$ & $F^{a}_{2\pi K}$ & $F^{e}_{K^* L}$  & $F^{e}_{K^* \parallel}$ & $F^{e}_{K^* \perp}$ \\
\hline
 $(0.39+i8.16)10^{-4}$  & $(1.99-i 3.36)10^{-2}$ & 0.36 & 0.06 & 0.11 \\ \hline \hline
 $F^{a}_{1K^* \phi L}$ & $F^{a}_{1K^* \phi \parallel}$ & $F^{a}_{1K^* \phi \perp}$ &
$F^{a}_{2K^* \phi L}$ & $F^{a}_{2K^* \phi \parallel}$ \\ \hline
 $(-1.4-i 1.0)10^{-3}$ & $(6.6+i6.5)10^{-4}$ & $(-1.2-i6.4)10^{-3}$ & $-0.03+i0.14$ &
 $0.03-i0.02$\\ \hline \hline
  $F^{a}_{2K^* \phi \perp}$ & $F^{e}_{\rho L}$ & $F^{e}_{\rho \parallel}$  & $F^{e}_{\rho \perp}$
   & $F^{a}_{1 \rho K^* L}$  \\\hline
  $0.06-i0.11$ &  0.31 & 0.04 & 0.08 & $(-2.3-i5.4)10^{-3}$ \\ \hline \hline
  $F^{a}_{1\rho K^* \parallel}$ & $F^{a}_{1\rho K^* \perp }$ & $F^{a}_{2 \rho K^* L}$ & $F^{a}_{2 \rho K^* \parallel}$ & $F^{a}_{2\rho K^* \perp}$
   \\\hline
 $(3.1+i0.9)10^{-4}$ & $(-1.9+i2.9)10^{-3}$ & $0.03+i0.16$ & $(0.65-i8.3)10^{-2}$ & $0.01-i0.17$\\
   \end{tabular}
\end{ruledtabular}
\end{table}
 From the table, we clearly see that the
annihilation contributions from $(V-A)\otimes  (V-A)$ operators
which correspond to $F^{a}_{1PP}$ and $F^{a}_{1VV}$ are negligible.
To be more clear understanding the results, we use $B\to PP$ decays
to illustrate the property. For $B\to PP$ decays, the factorized
amplitude associated with the $(V-A)\otimes (V-A)$ interaction for
annihilated topology can be expressed as \cite{CGHW}
\begin{eqnarray}
\langle P_{1} P_{2} | \bar{q}_{1}\gamma^{\mu}(1-\gamma_5) q_{2}\,
\bar{q}_{3} \gamma^{\mu} (1-\gamma_5) b | \bar{B}
\rangle_{a}=-if_{B} (m^{2}_{1}-m^{2}_{2})F^{P_1
P_2}_{0}(m^{2}_{B})\label{eq:anni1}
\end{eqnarray}
where $m_{1(2)}$ are the masses of outgoing particles and $F^{P_1
P_2}_{0}(m^2_{B})$ corresponds to the time-like form factor, defined
by
 \begin{eqnarray}
 \langle 0| \bar{q} \gamma^{\mu} \gamma_{5} b| \bar{B}(p_B)
\rangle &=&i f_{B} p^{\mu}_{B}\,,
\nonumber\\
 \langle P_{1}(p_1) P_{2}(p_2) | \bar{q}_{1} \gamma_{\mu} q_{2} |
 0\rangle & =& \left[q_{\mu}-\frac{m^{2}_1 -m^{2}_{2}}{Q^2}Q_{\mu}
 \right] F^{P_1 P_2}_{1}(Q^2) + \frac{m^2_1-m^2_2}{Q^2}Q_{\mu} F^{P_1
 P_2}_{0}(Q^2)\,,
\end{eqnarray}
respectively, with $q=p_1-p_2$ and $Q=p_1+p_2$. From Eq.
(\ref{eq:anni1}), it is clear that if $m_{1}= m_{2}$, the factorized
effects of annihilation topology vanish. However, the cancelation
factor will be removed when the interactions correspond to
$(V-A)\otimes (V+A)$ operators \cite{CGHW}.

\subsection{Experimental inputs and predictions of the SM}

As mentioned before, the accuracies of some experimental data on BRs
and CPAs are quite well, thus we could utilize these observed values
to constrain the new parameters of the $Z^{\prime}$ model. To be
more clear to know what the experimental inputs and the predictions
are, in the following we definitely display the ranges of current
experimental data for the inputs. Hence, taking the world averages
with $2\sigma$ errors presented in Ref. \cite{HFAG}, the inputs of
BRs are $B_{d}\to K^{0} \phi$, $B\to K^{*0} \phi$ and all $B\to\pi K
$ decays, and their limits are taken to be
\begin{eqnarray}
 &&7.3  < BR(B_{d}\to K \phi)10^{6}< 9.5, \ \ \,
 8.6  < BR(B_{d}\to K^{*} \phi)10^{6}< 10.4,
 \nonumber \\
 && 21.5<BR(B^{\pm}\to \pi^{\pm} K)10^{6} <26.7, \ \ \,  16.6<BR(B_d\to \pi^{\mp} K^{\pm})10^{6}
 <19.8
 \nonumber \\
 && 9.5 <BR(B_{d}\to \pi^{0} K)10^{6} <13.5 , \ \ \,  10.5<BR(B^{\pm}\to \pi^{0} K^{\pm})10^{6}
 <13.7. \label{eq:limit_br}
\end{eqnarray}
 Moreover, we also take into account the ratios of BRs, defined by
 \begin{eqnarray}
  R_{1}=\frac{\tau_{B^+}}{\tau_{B_d}} \frac{BR(B_{d}\to \pi^{\mp} K^{\pm})}{BR(B^{\pm} \to \pi^{\pm} K
  )}, \ \ \
        R_{c}= \frac{2BR(B^{\pm}\to \pi^{0} K^{\pm})}{BR(B^{\pm} \to \pi^{\pm} K
  )}, \ \ \
       R_{n}= \frac{BR(B_{d}\to \pi^{\mp} K^{\pm})}{2BR(B_{d} \to \pi^{0} K
  )},\label{eq:ratios}
   \end{eqnarray}
as $0.76<R_{1}<0.88$, $0.91<R_{c}<1.09$ and $0.74<R_{n}<0.88$
\cite{KOY_PRD72}.
 Since there are no measurements on the CPAs of $B\to K^{(*)}
\phi$, we artificially  set the limits as $0<|A_{CP}(B\to K^{(*)}
\phi)|<0.05$ in which the CPAs vanish in the SM. Other limits from
data are taken as $0<|A_{CP}(B^{+}\to \pi^{+} K^{0})| <5\%$,
$7.1\%<|A_{CP}(B_d\to \pi^{-} K^{+})| <14.7\%$, and
$0<|A_{CP}(B^{+}\to \pi^{0} K^{+})| <10\%$. Because there is no any
significant information on the CPA of $B_{d}\to \pi^{0} K^{0}$, we
leave the value as our prediction. In addition, we also take the
longitudinal polarization $R_{L}$ of $B_{d}\to K^{*0} \phi $ as the
input and the limit is chosen to be $ 44\%<R_{L}(B_{d}\to K^{*0}
\phi)< 57\%$.

\begin{figure}[htbp]
\includegraphics*[width=4.5 in]{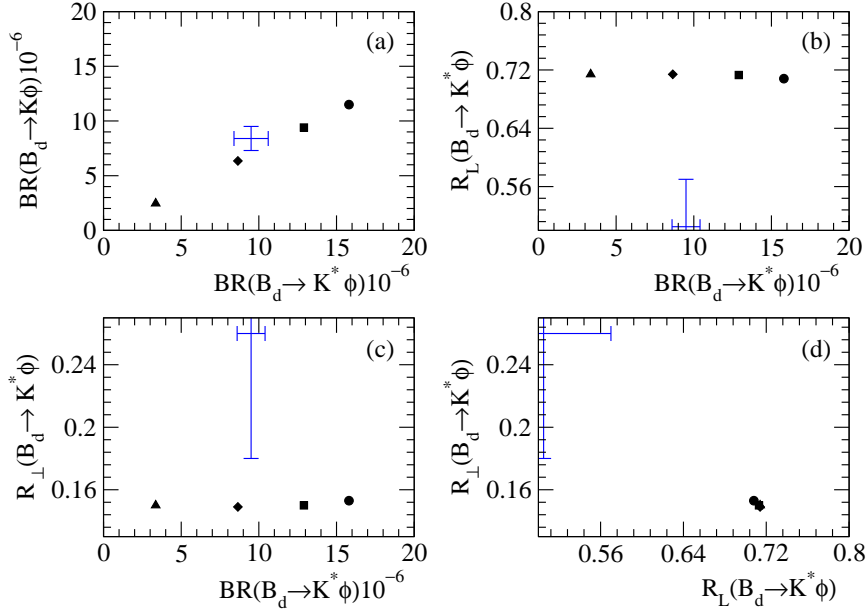}
\caption{ The SM predictions for  $B_d\to K^{(*)} \phi$ decays,
where (a), (b), (c) and (d) represent the correlations
%relationship of BR
between BRs of $K\phi$ and $K^* \phi$ modes,
%(b) and (c) are the correlations between
$R_{L(\perp)}$ and $BR(B_d\to K^* \phi)$,
 %and (c) stands for the correlation between
 and $R_{\perp}$ and $R_{L}$, respectively.
 The circle, square, diamond
and triangle-up symbols stand for the results of $\mu=1.3$, $1.5$,
$2.0$ and $4.0$ GeVs, respectively.  The error bars
%presented in the figures
are the world averages with $2\sigma$ errors. }
 \label{fig:phiks_sm}
\end{figure}

\begin{figure}[htbp]
\includegraphics*[width=4.5 in]{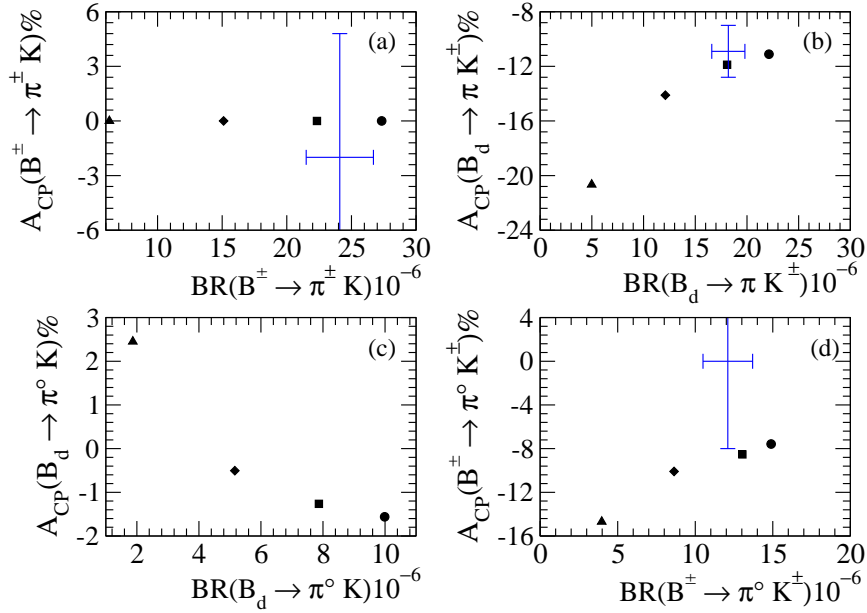}
\caption{  The SM predictions for the CPAs (in units of $10^{-2}$)
versus the corresponding BRs (in units of $10^{-6}$).
%, where the circle,
%square, diamond and triangle-up symbols stand for the results of
%$\mu=1.3$, $1.5$, $2.0$ and $4.0$ GeVs, respectively. The displayed
%error bars are the world
%averages with $2\sigma$ errors.
Legend is the same as Fig.~\ref{fig:phiks_sm}.
}
 \label{fig:pik_sm}
\end{figure}

Before going on discussing the contributions of the $Z^{\prime}$
model, it is worth knowing the SM results which are based on the
taken values in Sec.~\ref{sec:tinput}. Hence, the SM predictions on
BRs are
\begin{eqnarray}
BR(B_d\to K \phi)&=&8.98\times 10^{-6}, \ \ \ BR(B_{d}\to
K^{*}\phi)=12.9 \times 10^{-6}, \nonumber \\
BR(B^{\pm}\to \rho^{\pm} K^{*})&=&15.3 \times 10^{-6}, \ \ \
BR(B_{d}\to \rho^{\mp}
K^{*\pm})=13.4 \times 10^{-6}, \nonumber \\
BR(B^{\pm}\to \pi^{\pm} K)&=&22.2\times 10^{-6}, \ \ \
BR(B_d\to\pi^{\mp} K^{\pm})=19.0 \times 10^{-6},\nonumber \\
BR(B_{d}\to \pi^{0} K)&=&7.87\times 10^{-6}, \ \ \ BR(B^{\pm}\to
\pi^{0} K^{\pm})=12.5\times 10^{-6};
\end{eqnarray}
the ratios of BRs are estimated by $R^{SM}_{1}=0.92$,
$R^{SM}_{c}=1.18$ and $R^{SM}_{n}=1.22$; and the predictions on CPAs
are
\begin{eqnarray}
A_{CP}(B_{d}\to \rho^{\mp} K^{*\pm})&=& 22.0\%, \ \ \
A_{CP}(B_d\to\pi^{\mp}
K^{\pm})=-12.1\% ,  \nonumber \\
  A_{CP}(B_{d}\to \pi^{0} K)&=&-1.35\% , \ \ \ A_{CP}(B^{\pm}\to
\pi^{0} K^{\pm})=-8.3\%,
\end{eqnarray}
where the vanished CPAs are not shown. Moreover, the estimations of
the various PFs for $VV$ modes are also given to be
\begin{eqnarray}
R_{L}(B_d\to K^{*0}\phi )&=& 0.71, \ \ \ R_{\perp}(B_d\to
K^{*0}\phi)= 0.15,
 \nonumber \\
  R_{L}(B^{+}\to \rho^{+} K^{*0} )&=& 0.72, \ \ \ R_{\perp}(B^{+}\to \rho^{+} K^{*0})= 0.13,
 \nonumber \\
  R_{L}(B_d\to \rho^{-} K^{*+})&=& 0.52, \ \ \ R_{\perp}(B_d\to \rho^{-}
  K^{*+})=0.22.
\end{eqnarray}
According to our estimations, we see that compared to the data,
$B_d\to K^{*}\phi (B_{d}\to \pi^{0} K )$ has larger (smaller) BR,
the ratios $R_{1,c,n}$ don't fit the data well, and $R_{L}$ of
$B_{d} \to K^{*0} \phi$ is much larger than observations. We also
find $R_{L}(B_{d}\to \rho^{-} K^{*+})$ could be around $50\%$. For
displaying the influence of different scales, in
Fig.~\ref{fig:phiks_sm}, we present the correlations  between BRs in
$K\phi$ and $K^* \phi$ modes,
% the relationships of
$R_{L,\perp}$ and $BR(B_d\to K^* \phi)$, and
%the association of
$R_{L}$ and $R_{\perp}$, where the circle, square, diamond and
triangle-up symbols stand for the results of $\mu=1.3$, $1.5$, $2.0$
and $4.0$ GeVs, respectively. The error bars presented in the
figures are the world averages with $2\sigma$ errors. Similarly, we
also show the SM predictions on the CPAs of $B\to \pi K$ and the
corresponding BRs in Fig.~\ref{fig:pik_sm}.
%%%%%%%%%%%%%%%%%%%%%%%%%%%%%%%%%%%%%%%%%%%%%%%%%%%%%%%%%%%%%%%%%%%%%%%%%
%%%%%%%%%%%%%%%%%%%%%%%%%%%%%%%%%%%%%%%%%%%%%%%%%%%%%%%%%%%%%%%%%%%%%%%%
%%%%%%%%%%%%%%%%%%%%%%%%%%%%%%%%%%%%%%%%%%%%%%%%%%%%%%%%%%%%%%%%%%%%%%%%%
%%%%%%%%%%%%%%%%%%%%%%%%%%%%%%%%%%%%%%%%%%%%%%%%%%%%%%%%%%%%%%%%%%%%%%%%

\subsection{Results of the $Z^{\prime}$ model on $B\to \phi K^*$, $B\to \pi K$ and $B\to \rho
K^*$ decays}

Before performing the numerical calculations, we first discuss the
allowed regions of new effects which are from $\Delta
C^{\prime}_{9[7]}$. According to the results of
Refs.~\cite{BCLL_PLB580,BCLL_PLB598}, it is known that the unknown
parameters, defined by
$\xi^{LX}=(g_{2}m_{Z}/g_{1}m_{Z^{\prime}})^2B^{L*}_{sb}B^{X}_{dd}/V^*_{tb}V_{ts}$
with $X=L,\;R$, have been limited to be $|\xi^{LX}|\leq 0.02$ in
which $m^{\prime}_{Z}$ is at TeV-scale. That is, if we assume that
$B^{L}_{sb}\sim B^{R}_{sb}$ and $B^{R[L]}_{DD}\sim B^{L[R]}_{DD}$,
consequently we obtain $|\Delta C^{(\prime)}_{9[7]}|\leq 0.08$. In
the following analysis, we will take this value as the upper bound
of new effects. Since $\Delta C^{(\prime)}_{9[7]}$ in general are
complex, we totally have eight parameters for each $b\to s q\bar{q}$
with $q=(u,\;d)$ and $b\to s s\bar{s}$ decays. In principle, the
eight free parameters for $b\to s q\bar{q}$ could be fixed by the
eight chosen measurements such as four BRs and four CPAs in $B\to\pi
K$ decays. Then, using the constrained parameters we can make
predictions on $B\to \rho K^{*}$. Although there are no eight
measurements related to $b\to s s \bar{s}$ directly, however, due to
the new effects in Eq.~(\ref{eq:phik_xi}) for $B\to K \phi$ being
different from that in Eq.~(\ref{eq:phiks_xi}) for $\xi_{1(2)L}$ of
$B\to K^* \phi$, we find that when the data of $K \phi$ and $K^*
\phi$ are considered simultaneously, the unknowns has been strictly
constrained. For convenience, we parameterize the unknowns to be
$\Delta C_{9}=\eta_{LL} e^{i\phi_{LL}}$, $\Delta C_{7}=\eta_{LR}
e^{i\phi_{LR}}$, $\Delta C^{\prime}_{9}=\eta_{RL} e^{i\phi_{RL}}$,
$\Delta C^{\prime}_{7}=\eta_{RR} e^{i\phi_{RR}}$ so that
$|\eta_{XY}|\leq 0.08$ and $0\leq\phi_{XY}\leq 2\pi$ with $X$ and
$Y$ each being $L$ or $R$.

\begin{figure}[htbp]
\includegraphics*[width=4.5 in]{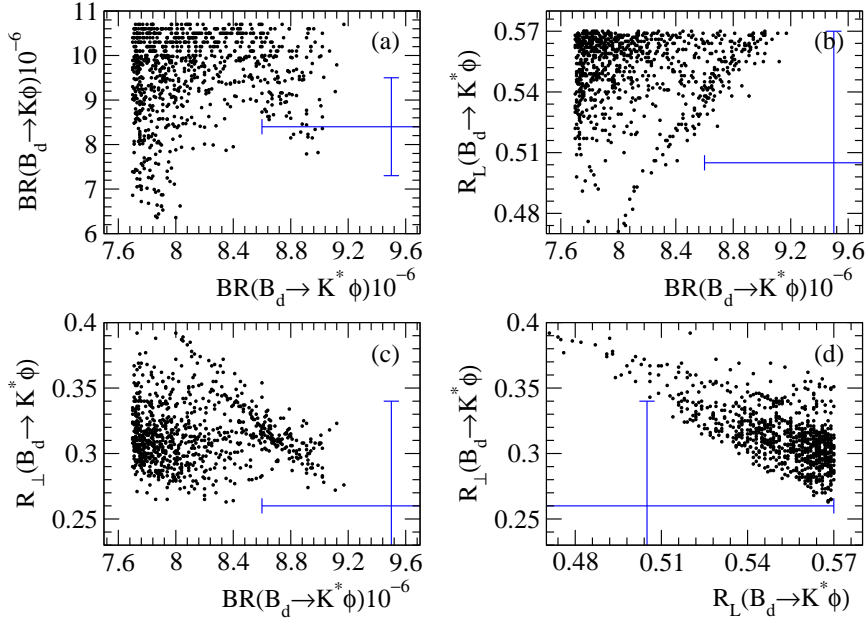}
\caption{(a), (b), (c) and (d) denote the correlations between BRs
of $K\phi$ and $K^* \phi$ modes, $R_{L(\perp)}$ and $BR(B_{d}\to K^*
\phi)$, and $R_{\perp}$ and $R_{L}$, respectively. The world
averages with $2\sigma$ errors are presented.}
 \label{fig:phiks}
\end{figure}

\begin{figure}[htbp]
\includegraphics*[width=4.5 in]{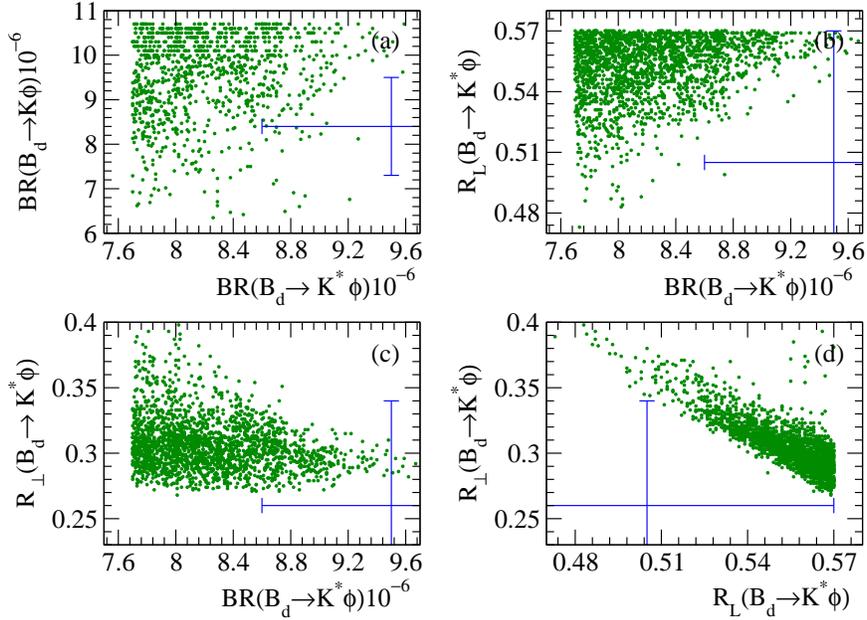}
\caption{  Legend is the same as Fig.~\ref{fig:phiks} but for
$\mu=1.3$ GeV.}
 \label{fig:phiks_s13}
\end{figure}

Now, we could investigate the contributions of the $Z^{\prime}$
model to the considering processes. At first, we study the decays
governed by $b\to s s\bar{s}$. It has been known that in terms of
flavor diagrams of Fig.~\ref{fig:flavor}, all effects contributing
to $B_d\to K \phi$ will also influence on $B_d\to K^* \phi$. It
could be expected that in SM-like models, to reduce the longitudinal
polarization $R_{L}$ of $B_d\to K^* \phi$ will also lower the BR of
$B\to K \phi$. We find that based on the hadronic values of
Table~\ref{tab:hardfun}, if we tune $\Delta C^{\prime}_{9[7]}=0$,
there are no solutions for the $\Delta C_{9[7]}$ to satisfy the data
of $BR(B_d\to K^{(*)} \phi)$ and $R_{L}(B_{d}\to K^* \phi)$ at the
same time. And also, if we set $\Delta C_{9[7]}=\Delta
C^{\prime}_{9[7]}$ or $\Delta C_{9[7]}=\Delta C^{\prime}_{7[9]}$
etc, no possible solutions are found. That is, in order to fit the
current experimental data, $\eta_{XY} (\phi_{XY})$ cannot have
simple relationship for different $X$ and $Y$. Hence, by taking each
$\eta_{XY}\leq 0.08$ and each $\phi_{XY}=[-\pi,\pi]$ and including
the limits of Eq.~(\ref{eq:limit_br}) and $44\%<R_{L}(B_{d}\to
K^{*0} \phi)< 57\%$, we present the possible solutions in
Fig.~\ref{fig:phiks}. By Fig.~\ref{fig:phiks}(a), we could see the
correlation of BR between $K\phi$ and $K^* \phi$. From the
Fig.~\ref{fig:phiks}(b) and (c), we see clearly how the changes of
$R_{L(\perp)}$ are associated with the BR of $K^*\phi$. We also
present the correlation of $R_{L}$ and $R_{\perp}$ in
Fig.~\ref{fig:phiks}(d). According to these results, it could be
concluded that $Z^{\prime}$ model which provides the left- and
right-handed couplings could solve the anomalies of small
$R_{L}(B\to K^* \phi)$. In addition, the $Z^{\prime}$ model also
provides the room for large $R_{\perp}(B\to K^* \phi)$, say above
$25\%$, in which $R_{\perp}$ of the SM is around $16\%$.
%%%%%%%%%%%%%%%%%%%%%%%%%%%%%%%%%%%%%%%%%%%%%%%%%%%%%%%%%%%%%%%%%%%%%%%%%
%%%%%%%%%%%%%%%%%%%%%%%%%%%%%%%%%%%%%%%%%%%%%%%%%%%%%%%%%%%%%%%%%%%%%%%%
As comparisons, we also show the results of $\mu=1.3$ GeV in
 Fig.~\ref{fig:phiks_s13}. Clearly, more solutions are allowed. We
 note that no solution can be found
 when $\mu > 1.5$
 GeV.
%%%%%%%%%%%%%%%%%%%%%%%%%%%%%%%%%%%%%%%%%%%%%%%%%%%%%%%%%%%%%%%%%%%%%%%%%
%%%%%%%%%%%%%%%%%%%%%%%%%%%%%%%%%%%%%%%%%%%%%%%%%%%%%%%%%%%%%%%%%%%%%%%%

\begin{figure}[htbp]
\includegraphics*[width=4.5 in]{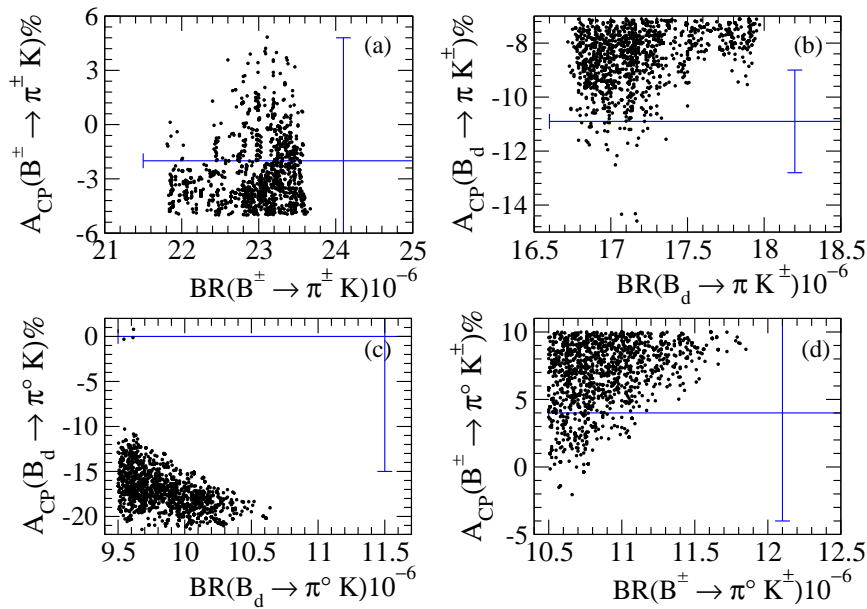}
\caption{The CPAs (in units of $10^{-2}$) versus the corresponding
BRs (in units of $10^{-6}$). The world averages with $2\sigma$
errors are included.}
 \label{fig:br-cp}
\end{figure}

\begin{figure}[htbp]
\includegraphics*[width=4.5 in]{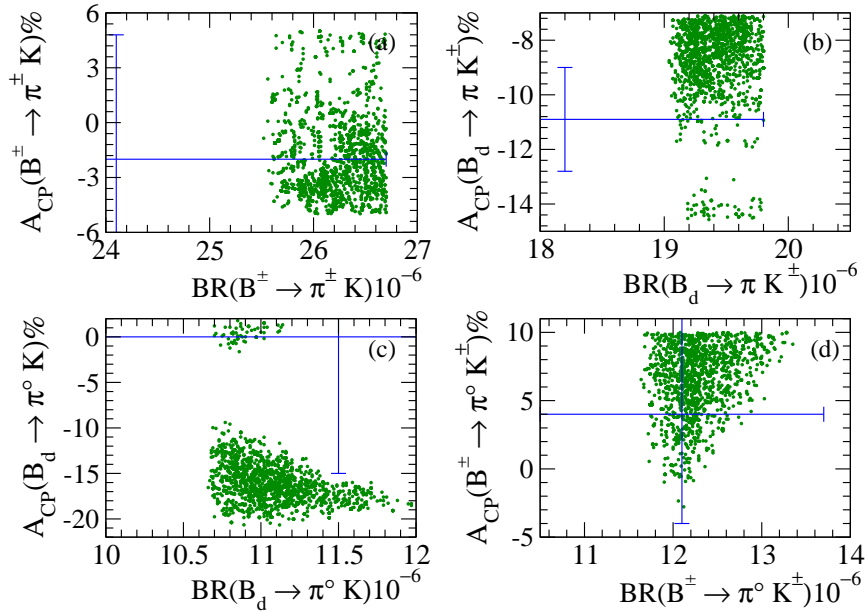}
\caption{Legend is the same as Fig.~\ref{fig:br-cp} but for
$\mu=1.3$ GeV.}
 \label{fig:br-cp_s13}
\end{figure}

Concerning the processes dictated by the decays $b\to s q\bar{q}$,
we also find that if we tune $\Delta C^{\prime}_{9[7]}=0$, or
$\Delta C_{9[7]}=\Delta C^{\prime}_{9[7]}$, or $\Delta
C_{9[7]}=\Delta C^{\prime}_{7[9]}$ etc, no possible solutions for
simultaneously matching the data are found. Hence, all unknowns
should be regarded as independent of parameters. Following the
formulas introduced in Sec.~\ref{sec:bpik}, the constraints of Eqs.
(\ref{eq:limit_br}) and (\ref{eq:ratios}) as well as the bounds of
CPA, we display the results in Fig.~\ref{fig:br-cp}. From the
figure, we could see that the CPA of $B^{\pm}\to \pi^{\pm} K$ could
be as large as $4\%$ while it vanishes in the SM. To satisfy all
current experimental data, the CPA of $B_{d}\to \pi^{0} K$ should be
smaller than $-10\%$. Since the CPA of $\pi^{\mp}K^{\pm}$ has a good
accurate measurement, if one could further confirm that the
magnitude of CPA for $B^{\pm}\to \pi^{\pm}K$ is small, say less than
$4\%$, we could conclude that the large CPA of $B_{d}\to \pi^{0} K$
could be a very good evidence to display the existence of new
physics, where the SM prediction is only around $-3\%$. Furthermore,
by Fig.~\ref{fig:br-cp}(d), we also see that the CPA of $B^{\pm}\to
\pi^{0} K^{\pm}$ could be much smaller than that of $B_{d}\to
\pi^{\mp} K^{\pm}$, in which they should have similar values in the
SM. And also, the results show that the CPAs of $\pi^{\mp} K^{\pm}$
and $\pi^{0} K^{\pm}$ favor to be opposite in sign but the SM
predicts the same sign. As mentioned in the end of
Sec.~\ref{sec:bpik}, when the decay amplitudes for
$B^{\pm}\to\pi^{\pm}K$, $B_{d}\to\pi^{\mp}K^{\pm}$ and
$B_d\to\pi^{0} K$ decays are determined, those for $B^{\pm}\to
\pi^{0} K^{\pm}$ decays are also fixed. Therefore, the sign
difference could be also as the clear evidence that new physics
exists.
%%%%%%%%%%%%%%%%%%%%%%%%%%%%%%%%%%%%%%%%%%%%%%%%%%%%%%%%%%%%%%%%%%%%%%%%%
%%%%%%%%%%%%%%%%%%%%%%%%%%%%%%%%%%%%%%%%%%%%%%%%%%%%%%%%%%%%%%%%%%%%%%%%
In Fig.~\ref{fig:br-cp_s13}, we also presented the results with
$\mu=1.3$ GeV. Since the data of $B\to \pi K$ have better
accuracies, we find that no possible solution appears when the scale
is smaller (larger) than $1.3(1.5)$ GeV.
%%%%%%%%%%%%%%%%%%%%%%%%%%%%%%%%%%%%%%%%%%%%%%%%%%%%%%%%%%%%%%%%%%%%%%%%%
%%%%%%%%%%%%%%%%%%%%%%%%%%%%%%%%%%%%%%%%%%%%%%%%%%%%%%%%%%%%%%%%%%%%%%%%

\begin{figure}[htbp]
\includegraphics*[width=4.5 in]{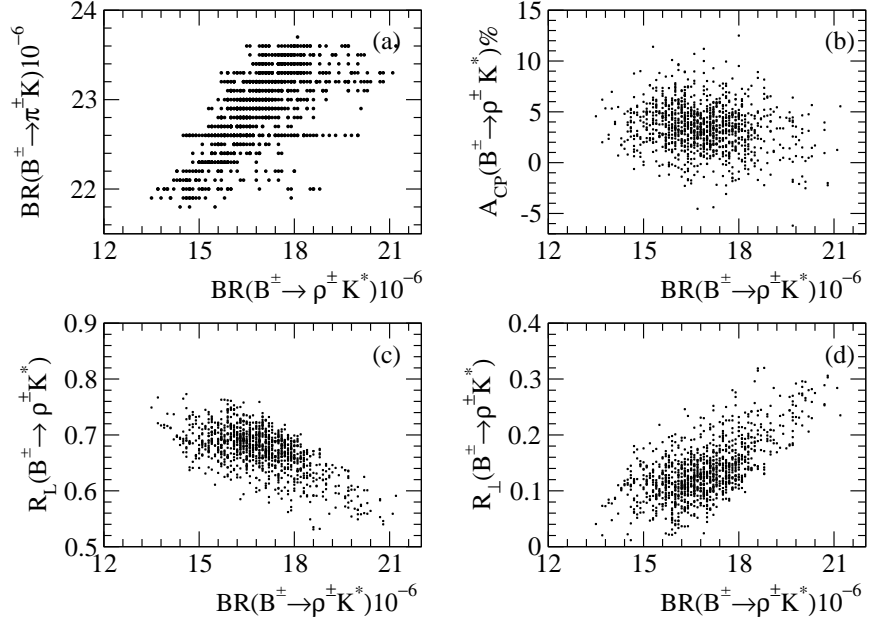}
\caption{(a) correlation of BR (in units of $10^{-6}$) between
$\rho^{\pm} K^*$ and $\pi^{\pm} K$; (b), (c) and (d) denote the
correlations between (CPA, $R_{\parallel}$, $R_{\perp}$) and the BR,
respectively.}
 \label{fig:crhonks}
\end{figure}

\begin{figure}[htbp]
\includegraphics*[width=4.5 in]{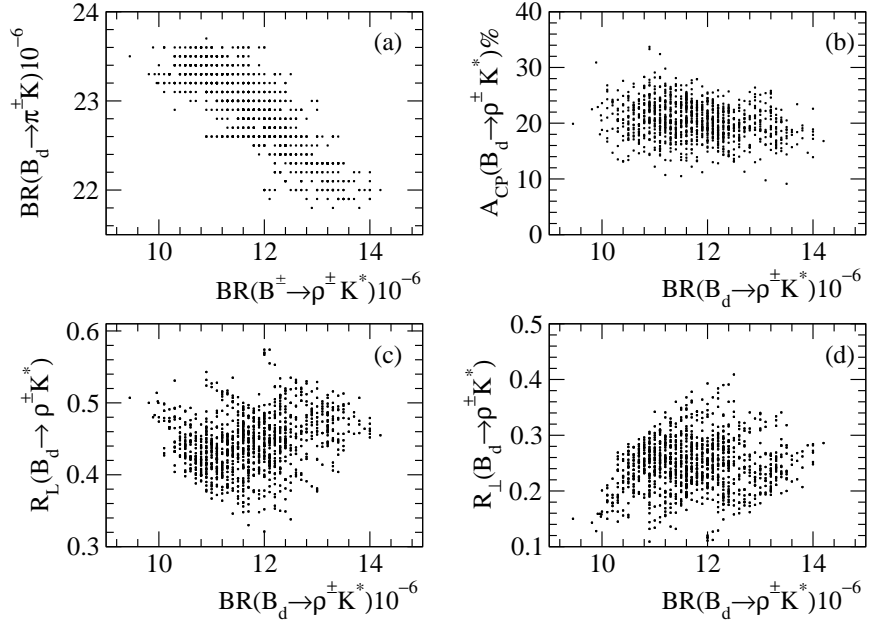}
\caption{The legend is the same as Fig.~\ref{fig:crhonks} but for
$B_{d}\to \rho^{\mp} K^{*\pm}$.}
 \label{fig:crhocks}
\end{figure}

Finally, we discuss the contributions of the $Z^{\prime}$ model to
the decays $B\to \rho K^*$. By the analysis in
Sec.~\ref{sec:brhoks}, we know that except the transverse parts, the
weak WCs for longitudinal polarization $R_{L}$ of $B\to \rho K^*$
should be the same for the decays $B\to \pi K$, i.e. they have the
same weak effective WCs $\zeta_{1,2}$, as shown in Eqs.
(\ref{eq:zeta}) and (\ref{eq:zeta_rhoks}). This is because the final
states in both processes have the same parity properties. However,
the case encountered in the decays $B\to K^{(*)}\phi$ is different
because the parity properties of final state $K\phi$ are different.
Hence, the values constrained by $B\to \pi K$ could directly make
predictions on $B\to \rho K^*$. We present the results of
$B^{\pm}\to \rho^{\pm} K^*$ and $B_{d}\to \rho^{\mp} K^{*\pm} $ in
Figs.~\ref{fig:crhonks} and \ref{fig:crhocks}, respectively. Since
the observed BR of $B_{d}\to \pi^{\pm} K$ has reached a good
accuracy, in Fig.\ref{fig:crhonks}(a) and Fig.\ref{fig:crhocks}(a)
we show how the BRs are associated with BR($B^{\pm}\to \pi^{\pm}
K$). Moreover, we display the CPA, $R_{L}$ and $R_{\perp}$ versus
the corresponding $BR$ in (b), (c) and (d) diagrams of both figures,
respectively. We note that the observed BR($R_{L}$) of $B^{\pm}\to
\rho^{\pm} K^*$ by BABAR and BELLE are not consistent each other.
The former observes $17.0\pm2.9 \pm 2.0(0.79\pm 0.08\pm 0.04)$
\cite{brhoks_babar} while the latter is $8.9\pm1.7\pm
1.0(0.43\pm0.11^{+0.05}_{-0.02})$ \cite{brhoks_belle}. By the
Fig.~\ref{fig:crhonks}, we could see clearly that (1) $B^{\pm}\to
\rho^{\pm} K^*$ can have sizable CPA in which it vanishes in the SM;
(2) $R_{L}$ could be less than $0.60$ while the corresponding BR is
above $15\times 10^{-6}$; (3) the solutions of small $R_{\perp}$
exist, i.e. $R_{\parallel}>>R_{\perp}$ where the prediction of SM is
$R_{\parallel}\sim R_{\perp}$. As for the results of $B_d\to
\rho^{\mp} K^{*\pm}$ shown in Fig.~\ref{fig:crhocks}, due to just
like the case of $B_{d}\to \pi^{\mp} K^{\pm}$ which the results with
new effects are similar to the SM, we expect that the derivations
from the SM are not too much. Hence, we could summarize the
favorable ranges of BRs, CPAs, $R_{L}$, and $R_{\perp}$ for
$(B^{\pm}\to \rho^{\pm} K^{*},\, B_{d}\to \rho^{\mp} K^{*\pm})$ are
$(17.1\pm 3.9,\,10.0\pm2.0)\times 10^{-6}$, $(3\pm 5,\, 21\pm 7)\%$,
$(0.66\pm 0.10,\, 0.44\pm 0.08)$ and $(0.14\pm 0.10,\, 0.25\pm
0.09)$, respectively.
%%%%%%%%%%%%%%%%%%%%%%%%%%%%%%%%%%%%%%%%%%%%%%%%%%%%%%%%%%%%%%%%%%%%%%%%%
%%%%%%%%%%%%%%%%%%%%%%%%%%%%%%%%%%%%%%%%%%%%%%%%%%%%%%%%%%%%%%%%%%%%%%%%
%%%%%%%%%%%%%%%%%%%%%%%%%%%%%%%%%%%%%%%%%%%%%%%%%%%%%%%%%%%%%%%%%%%%%%%%%
%%%%%%%%%%%%%%%%%%%%%%%%%%%%%%%%%%%%%%%%%%%%%%%%%%%%%%%%%%%%%%%%%%%%%%%%

\section{Summary}

We have studied the effects of nonuniversal  $Z^{\prime}$ model on
the processes dictated by the $b\to s q \bar{q}$ decays with $q=u$,
d, and s. By using the PQCD approach, we calculate the needed
hadronic matrix elements. For $B\to K^{(*)} \phi$ decays, we find
that their BRs and the $R_{L}$ of $B_{d}\to K^{*} \phi$ have
provided strict constraints on the new parameters. After marching
the currents data, we find the $R_{\perp}$ of $B_{d}\to K^{*0}\phi$
favors to be larger than $25\%$. For $B\to \pi K$ decays, by
requiring that the magnitude of $A_{CP}(B^{\pm}\to \pi^{\pm} K)$ is
less than $5\%$ and all BRs satisfy the current observations, we
find that the magnitude of CPA of $B_{d}\to \pi^{0}K$ should be
larger than $10\%$ but sign is the same as SM. Meanwhile, the CPA of
$B^{\pm}\to \pi^{0} K^{\pm}$ could be as low as few percent which is
indicated by the current experiments. Moreover, we also obtain that
the CPA of $B^{\pm}\to \pi^{0} K^{\pm}$ is opposite in sign to the
SM.

In sum, to satisfy current data, the new left- and right-handed
couplings have to be included simultaneously. It is clear that the
FC $Z^{\prime}$ model provides the needed couplings naturally. With
more physical observations and accurate data by $B$ factories, we
could further examine the effects of nonuniversal
$Z^{\prime}$ model.\\

{\bf Acknowledgments}\\

We thank Prof. Chao Qiang Geng and Prof. Cheng-Wei Chiang for useful
discussions. This work is supported in part by the National Science
Council of R.O.C. under Grant \#s:NSC-94-2112-M-006-009.

\section{Appendix: Distribution amplitudes and decay amplitudes }

\subsection{Distribution amplitudes }\label{app:distramp}

We describe the spin structures of meson to be
\begin{eqnarray}
\langle P(p) | \bar{q}_{2\beta}(z) q_{1\alpha}(0) |0\rangle &=& -
\frac{i}{\sqrt{2N_{c}}} \int^{1}_{0} dx e^{ix p\cdot z} \left\{
\gamma_{5} \not p \Phi_{P} (x) + \gamma_{5} m^{0}_{P} \Phi^{p}_{P}
(x) \right. \nonumber \\
&& \left. + m^{0}_{P} \gamma_{5} (\not n_{+} \not n_{-} -1)
\Phi^{\sigma}_{P} \right\}_{\alpha \beta}
\end{eqnarray}
for pseudoscalar, where $p=(p^{+},0, 0_{\perp})$,
$n_{+}=(1,0,0_{\perp})$, and $n_{-}=(0,1,0_{\perp})$; and
\begin{eqnarray}
\langle V(p,\epsilon_L) | \bar q_{2\beta }(z) q_{1\alpha }(0) | 0
\rangle
\nonumber\\
& & \hspace{-20mm} = \frac{1}{\sqrt{2N_c}}\int^1_0dx e^{ixp\cdot
z} \left[ m_{V} \not\epsilon_L
 \Phi_{V}(x)+
\not\epsilon_L \not p
 \Phi_{V}^t(x)
+ m_{V}
 \Phi_{V}^s(x)
\right]_{\alpha\beta}, \nonumber \\
%%%%%%%%%%%%%%%%%%%%%%%%%%%%%%%%%
\langle V(p,\epsilon_T)|\bar{q}_{2\beta }(z)q_{1\alpha }(0)|0> &=&
\frac{1}{\sqrt{2N_c}}\int_0^1 dx e^{ixp\cdot z} \left\{ \not
\epsilon_T \left[\not p
\Phi_{V}^T (x) + m_{Vo} \Phi_{V}^v (x) \right] \nonumber\right.\\
&&\left.+ \frac{m_{V}}{p\cdot n_{-}}
 i\epsilon_{T\mu\nu\rho\sigma}\gamma_5\gamma^\mu \epsilon^\nu p^\rho
n^\sigma\Phi_{V}^a (x)\right\}_{\alpha \beta}
\end{eqnarray}
for vector meson. The notations $\Phi_{P}$ and $\Phi^{(T)}_{V}$
denote the twist-2 wave functions while $\Phi^{p,\sigma}_{P}$ and
$\Phi^{t,s,v,a}_{V}$ stand for the twist-3 wave functions of
pseudoscalar and vector meson, respectively. Their explicit
expressions could be found in Refs. \cite{DA}.

\subsection{Hard functions for $B\to P_2 P_1$ decays} \label{app:hardamp_pp}

In terms of the spin structures of mesons defined by Appendix
~\ref{app:distramp}, we write the factorizable amplitudes for $B\to
P$ transition form factors and $B\to PP$ annihilations as
\begin{eqnarray}
F^{e}_{1P} &=& 8\pi C_{F} M_{B}^{2} \int_{0}^{1}
dx_{1}dx_{3}\int_{0}^{\infty }b_{1}db_{1}b_{3}db_{3}\Phi _{B}\left(
x_{1},b_{1}\right)
\nonumber \\
&&\times \left\{ \left[\left(1+x_{3}\right) \Phi _{P}\left(
x_{3}\right) +r_{P} ( 1-2x_{3}) \left( \Phi _{P}^{p}\left(
x_{3}\right) +\Phi _{P}^{\sigma }\left( x_{3}\right) \right)
\right]E_{e}\left( t_{e}^{\left( 1\right) }\right) \right.
\nonumber \\
&&\times \left. h_{e}\left( x_{1},x_{3},b_{1},b_{3}\right) +
2r_{P}\Phi _{P}^{p}\left( x_{3}\right)  E_{e}\left( t_{e}^{\left(
2\right) }\right) h_{e}\left( x_{3},x_{1},b_{3},b_{1}\right)
\right\} \;,
\end{eqnarray}
%%%%%%%%%%%%%%%%%%%%%%%%%%%%%%%%%%%%%%%%%%%%%%%%%%%%%%%%%%%
\begin{eqnarray}
F^{e}_{2P}&=& 16\pi C_{F}M_{B}^{2}
r_{P}\int_{0}^{1}dx_{1}dx_{3}\int_{0}^{\infty
}b_{1}db_{1}b_{3}db_{3}\Phi _{B}\left( x_{1},b_{1}\right)
\nonumber \\
&&\times \left\{ \left[ \Phi _{P}\left( x_{3}\right) +r_{P} \left(
(2+x_{3}) \Phi _{P}^{p}\left( x_{3}\right) -x_2 \Phi _{P}^{\sigma
}\left( x_{3}\right) \right) \right]E_{e}\left( t_{e}^{\left(
1\right) }\right) \right.
\nonumber \\
&&\times \left.  h_{e}\left( x_{1},x_{3},b_{1},b_{3}\right)+
2r_{P}\Phi _{P}^{p}\left( x_{3}\right)  E_{e}\left( t_{e}^{\left(
2\right) }\right) h_{e}\left( x_{3},x_{1},b_{3},b_{1}\right)
\right\} \;,
\end{eqnarray}
%%%%%%%%%%%%%%%%%%%%%%%%%%%%%%%%%%%%%%%%%%%%%%%%%%%%%%%%%%%%%
\begin{eqnarray}
F^{a}_{1P_{2} P_{1}}&=& -8\pi
C_{F}M_{B}^{2}\int_{0}^{1}dx_{2}dx_{3}\int_{0}^{\infty
}b_{2}db_{2}b_{3}db_{3}
\nonumber \\
&&\times \left\{ \left[ x_{3} \Phi_{P_1} ( x_{2}) \Phi_{P_2}(
1-x_{3}) + 2r_{P_1} r_{P_2} \Phi _{P_1}^{p}( x_{2}) \left(
(1+x_{3})\Phi_{P_2}^{p}( 1-x_{3})\right. \right. \right.\nonumber
\\
&& \left. \left. +(1-x_{3})\Phi_{P_2}^{\sigma }( 1-x_{3})\right)
\right] E_{a}\left( t_{a}^{(1) }\right) h_{a}(
x_{2},x_{3},b_{2},b_{3})
\nonumber \\
&&-\left[ x_{2} \Phi_{P_1}( x_{2}) \Phi_{P_2}( 1-x_{3})
+2r_{P_1}r_{P_2} \Phi_{P_2}^{p}( 1-x_{3})
\left((1+x_{2})\Phi_{P_2}^{p}(x_{2}) \right.\right. \nonumber
\\
&& \left. \left. \left. -(1-x_{2})\Phi_{P_1}^{\sigma}( x_{2})
\right) \right] E_{a}\left( t_{a}^{( 2) }\right)
h_{a}(x_{3},x_{2},b_{3},b_{2}) \right\} \;,
\end{eqnarray}
%%%%%%%%%%%%%%%%%%%%%%%%%%%%%%%%%%%%%%%%%%%%%%%%
\begin{eqnarray}
F^{a}_{2P_{2} P_{1}}&=& 16\pi
C_{F}M_{B}^{2}\int_{0}^{1}dx_{2}dx_{3}\int_{0}^{\infty}
b_{2}db_{2}b_{3}db_{3}
\nonumber \\
&&\times \left\{ \left[  r_{P_2} x_{3} \Phi _{P_1}(x_{2}) \left(
\Phi _{P_2}^{p} ( 1-x_{3}) + \Phi_{P_2}^{\sigma }(1-x_{3}) \right)
+2r_{P_1 }\Phi^{p} _{P_1}( x_{2}) \Phi_{P_2} ( 1-x_{3}) \right]
\right.
\nonumber \\
&&\times E_{a}\left( t_{a}^{( 1) }\right) h_{a}(
x_{2},x_{3},b_{2},b_{3}) + \left[ x_{2} r_{P_1}
\left(\Phi_{P_1}^{p}( x_{2})-\Phi _{P_1}^{\sigma}( x_{2}) \right)
\Phi_{P_2}(
1-x_{3}) \right. \nonumber \\
&& \left. \left. + 2r_{P_2}\Phi_{P_1}(x_2)\Phi
_{P_2}^{p}(1-x_3)\right] E_{a}\left( t_{a}^{( 2) }\right)
h_{a}(x_{3},x_{2},b_{3},b_{2}) \right\}\;
\end{eqnarray}
with $m_{P_{(i)}}/m_{B}$, where the hard functions $h_{e(a)}$ are
given by
\begin{eqnarray}
h_{e}(x_{1},x_{3},b_{1},b_{3}) &=&K_{0}\left( \sqrt{x_{1}x_{3}}
m_{B}b_{1}\right)S_t(x_3)
\nonumber \\
&&\times \left[ \theta (b_{1}-b_{3})K_{0}\left( \sqrt{x_{3}}
m_{B}b_{1}\right) I_{0}\left( \sqrt{x_{3}}m_{B}b_{3}\right) \right.
\nonumber \\
&&\left. +\theta (b_{3}-b_{1})K_{0}\left(
\sqrt{x_{3}}m_{B}b_{3}\right) I_{0}\left(
\sqrt{x_{3}}m_{B}b_{1}\right) \right] \;,
\label{he} \\
h_{a}(x_{2},x_{3},b_{2},b_{3}) &=&\left( \frac{i\pi }{2}\right)^{2}
H_{0}^{(1)}\left( \sqrt{x_{2}x_{3}}m_{B}b_{2}\right)S_t(x_3)
\nonumber \\
&&\times \left[ \theta (b_{2}-b_{3})H_{0}^{(1)}\left( \sqrt{x_{3}}
m_{B}b_{2}\right) J_{0}\left( \sqrt{x_{3}}m_{B}b_{3}\right) \right.
\nonumber \\
&&\left. +\theta (b_{3}-b_{2})H_{0}^{(1)}\left( \sqrt{x_{3}}
m_{B}b_{3}\right) J_{0}\left( \sqrt{x_{3}}m_{B}b_{2}\right) \right]
\;. \label{ha}
\end{eqnarray}
The evolution factor $E_{e(a)}$ are defined as
\begin{eqnarray}
E_{e}\left( t\right) &=&\alpha _{s}\left( t\right) S_{B}\left(
t\right)S_{P}\left( t\right)\;,
\nonumber \\
E_{a}\left( t\right) &=&\alpha _{s}\left( t\right) S_{P_1}(t)
S_{P_2}(t) \; \label{Eea}
\end{eqnarray}
where $S_{M}(t)$ denote the Sudakov factor of M-meson, the explicit
expressions could be found in Ref. \cite{CKLPRD66} and the
references therein.

\subsection{Hard functions for $B\to P V$ decays} \label{app:hardamp_pv}

Similarly, the factorizable amplitudes for $B\to PV$ modes are given
to be
\begin{eqnarray}
F^{a}_{1P V}&=&8\pi
C_{F}M_{B}^{2}\int_{0}^{1}dx_{2}dx_{3}\int_{0}^{\infty}
b_{2}db_{2}b_{3}db_{3}
\nonumber \\
&& \times \left\{ \left[ x_{3} \Phi_{V} ( x_{2}) \Phi_{P}(
1-x_{3})+2r_P r_{V}\Phi _{V}^{s}( x_{2}) \left(
(1+x_{3})\Phi_{P}^{p}(1-x_{3}) \right. \right. \right. \nonumber
\\
&& \left. \left. +(1-x_{3})\Phi_{P}^{\sigma }( 1-x_{3}) \right)
\right]E_{a}(t_a^{(1)})h_{a}(x_{2},x_{3},b_{2},b_{3}) -\left[ x_{2}
\Phi_{V}( x_{2}) \Phi_{P}( 1-x_{3}) \right. \nonumber \\
&& \left. +2r_P r_{V} \left((1+x_{2})\Phi_{V}^{s}( x_{2})
-(1-x_{2})\Phi_{V}^{t}( x_{2}) \right) \Phi_{P}^{p}(1-x_{3}) \right]
\nonumber \\
& &\times E_{a}(t_a^{(2)}) h_{a}(x_{3},x_{2},b_{3},b_{2}) \bigg\}\;,
\end{eqnarray}
%%%%%%%%%%%%%%%%%%%%%%%%%%%%%%%%%%%%%%%%%%%%%%%%%%%%%%%%%%%%%%%%
\begin{eqnarray}
F^{a}_{2P V}&=&-16\pi
C_{F}M_{B}^{2}\int_{0}^{1}dx_{2}dx_{3}\int_{0}^{\infty}
b_{2}db_{2}b_{3}db_{3}
\nonumber \\
&&\times \left\{ \left[  r_{P} x_{3} \Phi _{V}(x_{2}) \left( \Phi
_{P}^{p} (1- x_{3}) + \Phi_{P}^{\sigma }(1-x_{3}) \right) +2r_{V
}\Phi _{V}^{s}( x_{2}) \Phi_{P} ( 1-x_{3}) \right]\right.
\nonumber \\
&& \times E_{a}(t_a^{(1)}) h_{a}(x_{2},x_{3},b_{2},b_{3}) +\left[
2r_{P}\Phi_{V}(x_2)\Phi _{P}^{p}(1-x_3)\right. \nonumber \\
&& \left. \left. + x_{2} r_{V} \left(\Phi_{V}^{s}( x_{2})-\Phi
_{V}^{t}( x_{2}) \right) \Phi_{P}( 1-x_{3})\right] E_{a}(t_a^{(2)})
h_{a}(x_{3},x_{2},b_{3},b_{2}) \right\}\;
\end{eqnarray}
with $r_{V}=m_{V}/m_{B}$.

\subsection{Hard functions for $B\to V_{2}V_{1}$ decays}\label{app:hardamp_vv}

The needed factorizable amplitudes for $VV$ modes are given by
\begin{eqnarray}
F^{e}_{V_{2}L} &=&8\pi
C_{F}M_{B}^{2}\int_{0}^{1}dx_{1}dx_{3}\int_{0}^{\infty}
b_{1}db_{1}b_{3}db_{3}\Phi _{B}( x_{1},b_{1})
\nonumber\\
&& \times \left\{ \left [(1+x_{3})\Phi_{V_2}(x_{3}) +
r_{V_2}(1-2x_{3})(\Phi^{t}_{V_2}(x_3)+\Phi^{s}_{V_2}(x_{3}))\right]
\right.
\nonumber\\
&& \times E^e(t^{(1)}_{e}) h_{e}(x_{1},x_{3},b_{1},b_{3})
\nonumber\\
&&\left.+2r_{V_2}\Phi^{s}_{V_2}(x_3)E^e(t^{(2)}_{e})
h_{e}(x_{3},x_{1},b_{3},b_{1}) \right\}\;,
\end{eqnarray}
%%%%%%%%%%%%%%%%%%%%%%%%%%%%%%
\begin{eqnarray}
F^e_{V_{2}N} &=& 8\pi
C_{F}M_{B}^{2}\int_{0}^{1}dx_{1}dx_{3}\int_{0}^{\infty}
b_{1}db_{1}b_{3}db_{3}\Phi _{B}(x_{1},b_{1})
\nonumber\\
&& \times r_{V_1} \left\{ [\Phi^{T}_{V_2}(x_3)
+2r_{V_2}\Phi^{v}_{V_2}(x_3)+r_{V_2}x_{3}
(\Phi^{v}_{V_2}(x_3)-\Phi^{a}_{V_2}(x_3))]\right.
\nonumber \\
&& \times E^e(t^{(1)}_{e}) h_{e}(x_{1},x_{3},b_{1},b_{3})
\nonumber \\
&&\left.+r_{V_2}[\Phi^{v}_{V_2}(x_3)+\Phi^{a}_{V_2}(x_3)]
E^e(t^{(2)}_{e}) h_{e}( x_{3},x_{1},b_{3},b_{1})\right\}\;,
\end{eqnarray}
%%%%%%%%%%%%%%%%%%%%%%%%%%%%%%%%
\begin{eqnarray}
F^e_{V_{2}T} &=&16\pi
C_{F}M_{B}^{2}\int_{0}^{1}dx_{1}dx_{3}\int_{0}^{\infty}
b_{1}db_{1}b_{3}db_{3}\Phi _{B}( x_{1},b_{1})
\nonumber\\
&&\times r_{V_1}\left\{[\Phi^{T}_{V_2}(x_3)
+2r_{V_2}\Phi^{a}_{V_2}(x_3)-r_{V_2}x_{3}
(\Phi^{v}_{V_2}(x_3)-\Phi^{a}_{V_2}(x_3))]\right.
\nonumber \\
&& \times E^e(t^{(1)}_{e}) h_{e}(x_{1},x_{3},b_{1},b_{3})
\nonumber \\
&&\left.+r_{V_2}[\Phi^{v}_{V_2}(x_3)+\Phi^{a}_{V_2}(x_3)]
E^e(t^{(2)}_{e})h_{e}( x_{3},x_{1},b_{3},b_{1})\right\}\;,
\end{eqnarray}
%%%%%%%%%%%%%%%%%%%%%%%%%%%%%%%%%%%%%%%%%%%%%%%%%%
\begin{eqnarray}
F^a_{1V_{2} V_{1}L} &=&8\pi
C_{F}M_{B}^{2}\int_{0}^{1}dx_{2}dx_{3}\int_{0}^{\infty}
b_{2}db_{2}b_{3}db_{3}
\nonumber \\
&& \times \left\{ \left[-x_3 \Phi_{V_1}(x_2)\Phi_{V_2}(1-x_3)
+2r_{V_1}r_{V_2}\Phi^{s}_{V_1}(x_2) ((1-x_{3})\Phi^{t}_{V_2}(1-x_3)
\right. \right.\nonumber
\\
&& \left. +(1+x_{3})\Phi^{s}_{V_2}(1-x_3))\right]
 E^a(t_{a}^{(1)})h_{a}(x_{2},x_{3},b_{2},b_{3})
\nonumber \\
&& + \left [ x_{2}\Phi_{V_1}(x_{2})\Phi_{V_2}(1-x_{3})
+2r_{V_1}r_{V_2}\Phi^{s}_{V_2}(1-x_{3}) ((1-x_{2})
\Phi^{t}_{V_1}(x_2) \right. \nonumber \\
&& \left. \left. - (1+x_{2}) \Phi^{s}_{V_1}(x_2))\right]
E^a(t_{a}^{(2)}) h_{a}(x_{3},x_{2},b_{3},b_{2}) \right\}\;,
\end{eqnarray}
%%%%%%%%%%%%%%%%%%%%%%%%%
\begin{eqnarray}
F^a_{1V_{2} V_{1}N}&=&-8\pi
C_{F}M_{B}^{2}\int_{0}^{1}dx_{2}dx_{3}\int_{0}^{\infty}
b_{2}db_{2}b_{3}db_{3}
\nonumber \\
&&\times r_{V_1}r_{V_2}\left\{
\left[(1+x_{3})(\Phi^{v}_{V_1}(x_2)\Phi^{v}_{V_2}(1-x_3)
+\Phi^{a}_{V_1}(x_2)\Phi^{a}_{V_2}(1-x_3))\right.\right.
\nonumber \\
&& \left.+ (1-x_3)(\Phi^{v}_{V_1}(x_2)\Phi^{a}_{V_2}(1-x_3)
+\Phi^{a}_{V_1}(x_2)\Phi^{v}_{V_2}(1-x_3))\right]
E^a(t_{a}^{(1)})h_{a}(x_{2},x_{3},b_{2},b_{3})
\nonumber\\
&& -\left[(1+x_{2})(\Phi^{v}_{V_1}(x_2)\Phi^{v}_{V_2}(1-x_3)
+\Phi^{a}_{V_1}(x_2)\Phi^{a}_{V_2}(1-x_3)) \right.
\nonumber \\
&&  \left.-(1-x_{2})(\Phi^{v}_{V_1}(x_2)\Phi^{a}_{V_2}(1-x_3)
+\Phi^{a}_{V_1}(x_2)\Phi^{v}_{V_2}(1-x_3))\right] \nonumber \\
&& \left. \times E^a(t_{a}^{(2)})h_{a}(x_{3},x_{2},b_{3},b_{2})
\right\}\;,
\end{eqnarray}
%%%%%%%%%%%%%%%%%%%%%%%%%%%%%%%%%%%%
\begin{eqnarray}
F^a_{1V_{2} V_{1}T} &=&-16\pi
C_{F}M_{B}^{2}\int_{0}^{1}dx_{2}dx_{3}\int_{0}^{\infty}
b_{2}db_{2}b_{3}db_{3}
\nonumber \\
&&\times r_{V_1}r_{V_2}\left\{
\left[(1-x_{3})(\Phi^{v}_{V_1}(x_2)\Phi^{v}_{V_2}(1-x_3)
+\Phi^{a}_{V_1}(x_2)\Phi^{a}_{V_2}(1-x_3))\right.\right.
\nonumber \\
&& \left.+(1+x_3)(\Phi^{v}_{V_1}(x_2)\Phi^{a}_{V_2}(1-x_3)
+\Phi^{a}_{V_1}(x_2)\Phi^{v}_{V_2}(1-x_3))\right]
E^a(t_{a}^{(1)})h_{a}(x_{2},x_{3},b_{2},b_{3})
\nonumber\\
&& +\left[(1-x_{2})(\Phi^{v}_{V_1}(x_2)\Phi^{v}_{V_2}(1-x_3)
+\Phi^{a}_{V_1}(x_2)\Phi^{a}_{V_2}(1-x_3)) \right.
\nonumber \\
&&  \left.-(1+x_{2})(\Phi^{v}_{V_1}(x_2)\Phi^{a}_{V_2}(1-x_3)
+\Phi^{a}_{V_1}(x_2)\Phi^{v}_{V_2}(1-x_3))\right] \nonumber
\\&& \left. \times E^a(t_{a}^{(2)})h_{a}(x_{3},x_{2},b_{3},b_{2})
\right\}\;,
\end{eqnarray}
%%%%%%%%%%%%%%%%%%%%%%%%%%%%%%%%%%%%%%%%%
\begin{eqnarray}
F^a_{2V_{2} V_{1}L} &=&16\pi
C_{F}M_{B}^{2}\int_{0}^{1}dx_{2}dx_{3}\int_{0}^{\infty}
b_{2}db_{2}b_{3}db_{3}
\nonumber \\
&& \times \left\{\left[r_{V_2} x_3 \Phi_{V_1}(x_2)
(\Phi^{t}_{V_2}(1-x_3)+\Phi^{s}_{V_2}(1-x_3))-2r_{V_1}\Phi^{s}_{V_1}(x_2)
\Phi_{V_2}(1-x_3) \right] \right.
\nonumber \\
&& \times E^a(t^{(1)}_{a})h_{a}(x_{2},x_{3},b_{2},b_{3})
\nonumber\\
&& + \left[r_{V_1} {x_2} (\Phi^{t}_{V_1}(x_2) - \Phi^{s}_{V_1}(x_2)
) \Phi_{V_2}(1-x_3)+ 2r_{V_2} \Phi_{V_1}(x_2)\Phi^{s}_{V_2}(1-x_3)
\right]
\nonumber \\
&& \left.\times E^a(t_{a}^{(2)})h_{a}(x_{3},x_{2},b_{3},b_{2})
\right\}\;,
\end{eqnarray}
%%%%%%%%%%%%%%%%%%%%%%%%%%%%%%%%%%%%%%%%%%
\begin{eqnarray}
F^a_{2V_{2} V_{1}N}&=&16\pi
C_{F}M_{B}^{2}\int_{0}^{1}dx_{2}dx_{3}\int_{0}^{\infty}
b_{2}db_{2}b_{3}db_{3}
\nonumber \\
&& \times\left\{r_{V_1}(\Phi^{v}_{V_1}(x_2)
+\Phi^a_{V_1}(x_{2}))\Phi^{T}_{V_2}(1-x_3)
E^a(t^{(1)}_{a})h_{a}(x_2,x_3,b_2,b_3)\right.
\nonumber \\
&&+ \left. r_{V_2}\Phi^{T}_{V_1}(x_2)(\Phi^{v}_{V_2}(1-x_3)
-\Phi^a_{V_1}(1-x_{3}))
E^a(t^{(1)}_{a})h_{a}(x_3,x_2,b_3,b_2)\right\}\;,
\end{eqnarray}
%%%%%%%%%%%%%%%%%%%%%%%%%%%%%%%%%%%%%%%%%
\begin{eqnarray}
F^a_{2V_{2} V_{1}T}&=&32\pi
C_{F}M_{B}^{2}\int_{0}^{1}dx_{2}dx_{3}\int_{0}^{\infty}
b_{2}db_{2}b_{3}db_{3}
\nonumber \\
&& \times\left\{r_{V_1}(\Phi^{v}_{V_1}(x_2)
+\Phi^a_{V_1}(x_{2}))\Phi^{T}_{V_2}(1-x_3)
E^a(t^{(1)}_{a})h_{a}(x_2,x_3,b_2,b_3)\right.
\nonumber \\
&&+ \left. r_{V_2}\Phi^{T}_{V_1}(x_2)(\Phi^{v}_{V_2}(1-x_3)
-\Phi^a_{V_2}(1-x_{3}))
E^a(t^{(2)}_{a})h_{a}(x_3,x_2,b_3,b_2)\right\}\;.
\end{eqnarray}
We define $r_{V_{i}}=m_{V_{i}}/m_{B}$.

\end{document}